\documentclass[aps, pra, preprint, superscriptaddress]{revtex4-1}

\usepackage{times,amsmath}
\usepackage{bm}
\usepackage{epsfig}
\usepackage{amsfonts}
\usepackage{subfigure}
\usepackage{color}
\usepackage{hyperref}
\usepackage[normalem]{ulem}
\usepackage{lipsum}
\usepackage[utf8]{inputenc}

\newcommand{\e}{\varepsilon}
\newcommand{\f}{\varphi}
\newcommand{\w}{\omega}

\renewcommand{\Re}{{\rm Re}}

\begin{document}

\title{Enhanced stability, bistability, and exceptional points in saturable active photonic couplers}

\author{Yertay Zhiyenbayev}
\email{yertay.zhiyenbayev@nu.edu.kz}
\affiliation{Department of Physics,\\ Nazarbayev University, KZ-010000, Kazakhstan}

\author{Yannis Kominis}
\email{gkomin@central.ntua.gr}
\affiliation{School of Applied Mathematical and Physical Science,\\ National Technical University of Athens, GR-10682, Greece}

\author{Constantinos Valagiannopoulos}
\email{konstantinos.valagiannopoulos@nu.edu.kz}
\affiliation{Department of Physics,\\ Nazarbayev University, KZ-010000, Kazakhstan}

\author{Vassilios Kovanis}
\affiliation{Bradley Department of Electrical and Computer Engineering,\\ Virginia Tech Research Center, VA-22203, Virginia, USA}
\affiliation{School of Optics and Photonics - CREOL, \\ The University of Central Florida, Orlando, Florida 32816-2700, USA}

\author{Anastasios Bountis}
\affiliation{Department of Mathematics,\\ Nazarbayev University, KZ-010000, Kazakhstan}

\begin{abstract}
A generic photonic coupler with active and lossy parts, gain saturation and asymmetric characteristics is examined. Saturable activity is shown to be able to enhance the overall stability of the steady states, prevent evolution to undesirable unbounded modes and allow for bistable operation in specific regions of parametric space. Both stability and bistability are studied in the phase space of the system, where the basins of attraction of each state are identified, providing an accurate description of the dependence of the electric fields on the initial conditions. Continuous families of exceptional points are detected via suitable regulation of the coupling and asymmetry features of the configuration. In this way, a complete description of the nonlinear dynamics landscape is provided, which should be crucial for multiple application-driven designs incorporating such a ubiquitous optical component.
\end{abstract}

\maketitle 

\section{Introduction}
The nonlinear coherent optical coupler \cite{Jensen_82, Daino_85} is one of the most fundamental elements for multiple key technological architectures and photonic integrated circuits \cite{Saleh}, allowing for applications related to power-dependent directed transport of energy \cite{Feng_13}, unidirectional propagation \cite{Feng_11} and optical isolation \cite{Jalas_13, Shi_15}. The possibility of engineering the gain and loss characteristics of such devices results in a rich set of non-Hermitian dynamical features that have no counterpart in conservative (Hermitian) configurations and has been analyzed for decades (see e.g. \cite{Thompson_86, Chen_92}). The generality of the underlying model, consisting of coupled-mode equations, suggests its applicability to a wide class of isomorphic non-Hermitian dimers, describing other photonic systems such as twisted fiber amplifiers \cite{Malomed_17} as well as quantum systems \cite{Kenkre_86, Dattoli_90}.

Most of the setups considered so far are structurally symmetric, with balanced loss and gain leading to Parity-Time (PT) symmetric configurations; indeed, in recent years there has been tremendous progress in the theoretical formulations and experimental implementations of such systems \cite{DawnNonHermitian}. The overall lossless nature of these layouts gives rise to optimal responses with large applicability potential spanning from single mode lasing \cite{SingleModeLaser, PTSymmetricMicrorings}, coherent perfect absorption \cite{Longhi_10} and optical switches with plasmonic waveguides \cite{PTWaveguides, MyJ73, PTPlasmonics,BenistySwitches} to invisible acoustic sensors \cite{InvisibleAcousticSensor}. Furthermore, interesting features and properties like tunable quantum phase transitions \cite{NonHermitianPhotonics} and whispering-gallery-mode resonances \cite{WhisperingGallery} were revealed due to PT symmetry. Most importantly, the PT-symmetric condition, when imposed on nonlinear structures, has led to significant results related to active control of light \cite{NonlinearSwitching}, controlled power transport \cite{Ramezani_10} and robust soliton propagation \cite{NonlinearWaves}.

However, under the symmetric presence of activity and dissipation, the system often evolves to states with unbounded electric field amplitudes, which is undesirable for realistic applications. As an example, two waveguides \cite{Ramezani_10} with identical wave propagation numbers and exactly opposite gain and loss coefficients support either an asymmetric unbounded state \cite{Barashenkov_13, Christodoulides_16} or a symmetric bounded nonlinear supermode, therefore not allowing for capabilities of directed power transfer between the two waveguides in a stable fashion. A remedy to such an unwanted behavior is the introduction of gain/loss asymmetry which, as has been recently shown, not only enhances the stability of the system \cite{Kominis_17}, but also admits controlled directed power transport enabled by the emergence of additional strongly asymmetric modes \cite{Kominis_16}. It is worth mentioning that the asymmetry as a stabilizer has been considered for other non-Hermitian photonic systems consisting of coupled lasers \cite{Kominis_17b, Choquette_17} but also for more general configurations of paired oscillators \cite{Roy_19}. In addition, investigations on the key role of asymmetry on the formation and propagation of self-localized beams in non-Hermitian configurations have shown that continuous families of solitary waves can be formed under generic conditions, not necessarily restricted by symmetry requirements \cite{Kominis_19b}; as a result, they exhibit a rich set of propagation features such as dynamical trapping and controllable routing once the gain and loss spatial distribution is properly engineered \cite{Kominis_15a, Kominis_15b}.

Apart from asymmetry in structure and excitation, another important characteristic determining the functionality of paired optical structures is saturation. This property expresses the difficulty of photonic matter to be receptive to fields of large magnitudes developed therein; accordingly, the texture of the material changes to prevent unbounded power increase within its volume. Such behavior also yields a nonlinear effect that is most commonly present in active media. Indeed, saturation causes significant damping influence on the instability of evanescently coupled arrays \cite{NonlinearGain}, acting as a natural stability booster. Moreover, saturable active matter, except for being more realistic, has been utilized for the study of optical cavity effects in nanowire lasers and waveguides \cite{OpticalCavityEffects} and ultrafast all-optical absorption switch for photonic crystals \cite{UltrafastPhotonicCrystal}.

In this work, we combine the aforementioned characteristics of asymmetry and saturation to study the dynamical features of continuous waves in a generic asymmetric active coupler with saturable gain. After analytically determining the fixed points corresponding to the Nonlinear Supermodes (NS) of the system, by linear stability analysis, we scan the parametric space and identify the response of the device. Due to the considered gain saturation, the stability of the system gets enhanced and under certain conditions bistability \cite{Thirstrup_95} emerges, which is the backbone of multiple memory \cite{FilmCoupledMetasurfaces} and filtering \cite{Ludge_18} applications. In cases where one or two stable NS exist we examine the influence of the initial conditions on the response of the coupler and, by numerically computing the evolution of the excited fields, we provide reliable basins of attraction \cite{Thompson} for the stable NSs and the undesirable unbounded state that govern the operation of the setup. Moreover, such information is crucial when designing components for the dynamic re-configuration between supported steady states \cite{PeriodicOrbitsBasins}, high-precision measurement and detection \cite{Roukes_07}. Finally, we search the non-Hermitian structure of our system for spectral degeneracies known as Exceptional Points (EP) \cite{Heiss} that are related to substantial sensitivity capabilities \cite{Christodoulides} as well as mode conversion utilities \cite{Fan_19}. We believe that the thorough and multi-faceted analysis of such a general layout, while interesting for its nonlinear dynamics content, is most important for its usefulness to a wealth of applications that are involved.

\section{Coupled-Mode equations and Nonlinear Supermodes}

\subsection{System Model and Coupled Mode equations}
\begin{figure}[ht!]
\centering
\includegraphics[width=7.2cm]{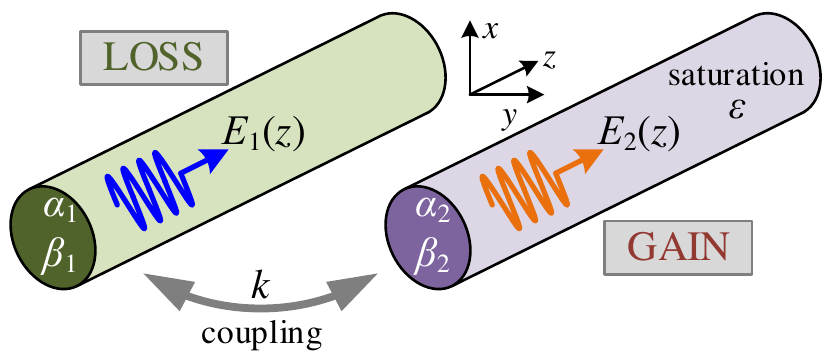}
\caption{Sketch of the examined setup. Two waveguides (one passive and one active) are evanescently coupled (by the constant $k$) and the developed waves $E_1(z),E_2(z)$ are propagating along axis $z$. The complex refractive index of the passive waveguide is denoted by $(\beta_1+i\alpha_1)$ and that of the active (also saturable with saturation constant $\e$) is $(\beta_2+i\alpha_2)$.}
\label{fig:Fig0}
\end{figure}

We consider a pair of parallel waveguides positioned along the $z$ axis of our Cartesian coordinate system $(x,y,z)$, depicted in Fig. \ref{fig:Fig0}. $E_j(z)$ with $j=1,2$ are the electric field complex phasors in each waveguide $j$ when dropping the transverse spatial dependence $xy$ and the vectorial nature of the modes. The harmonic time is of the form $\exp(-i\w t)$ and propagation along the $z>0$ axis is examined. The first waveguide ($j=1$) is lossy and, in its linear operation, characterized by a complex propagation constant $(\beta_1+i\alpha_1)$, while $\alpha_1>0$ measures the magnitude of the loss. The second waveguide ($j=2$) has gain; thus, $\alpha_2<0$ and, inevitably, its response is clipped by saturation with constant $\e>0$. Both waveguides are nonlinear and obeying the Kerr effect, namely their refractive indexes are proportional to the squared field magnitude with (common) proportionality constant $\gamma$. The two elements are evanescently coupled and their interaction is expressed through a positive linear coupling parameter $k>0$ (nonlinear coupling is ignored). \\


Wave propagation into the aforementioned system is described by the following Coupled-Mode Equations \cite{CMT}:
\begin{eqnarray}
-i\frac{d E_1}{d z}&=&\left(\beta_1+i\alpha_1\right) E_1+\gamma |E_1|^2 E_1+\frac{k}{2}E_2,
\label{CME1}\\
-i\frac{d E_2}{d z}\!&=&\!\left(\beta_2+i\frac{\alpha_2}{1+\e|E_2|^2}\right)\! E_2 \!+\! \gamma |E_2|^2 E_2 \!+\! \frac{k}{2}E_1. 
\label{CME2}
\end{eqnarray}
One may write the two complex fields in polar form: $E_j(z)=A_j(z)\exp[ibz+i\f_j(z)]$, with common propagation constant $b\in\mathbb{R}$, amplitudes $A_j>0$ and phases $\f_j\in\mathbb{R}$, for $j=1,2$. If so, the equations \eqref{CME1},\eqref{CME2} can be re-written as follows by separating real from imaginary parts:
\begin{eqnarray}
\dot{A}_1&=&-\alpha_1 A_1-\frac{k}{2}A_2\sin\f,
\label{ACME1}\\
\dot{A}_2&=&-\frac{\alpha_2}{1+\e A_2^2} A_2+\frac{k}{2}A_1\sin\f,
\label{ACME2}\\
\dot{\f}\!&=&\!(\beta_2\!-\!\beta_1)\!+\!\gamma \left(A_2^2\!-\!A_1^2\right)\!+\!\frac{k}{2}\left(\frac{A_1}{A_2}-\frac{A_2}{A_1}\right)\cos\f,
\label{ACME3}
\end{eqnarray}
where $\f=\f_2-\f_1$ is the phase difference between the two waves and a dot denotes the derivative with respect to $z$.

\subsection{Nonlinear Supermodes}
The Nonlinear Supermodes (NS) of the system correspond to steady states propagating with constant values of the field amplitudes and phase difference $\left\{A_1,A_2,\f\right\}$. They can be found as fixed points of the dynamical system defined by Eqs.	(\ref{ACME1})-(\ref{ACME3}) by solving the nonlinear algebraic system obtained after setting the derivatives equal to zero. The NS are completely described by four parameters, namely the amplitude of one of the waveguides ($A_1$), the ratio of the squared field amplitudes ($R=(A_2/A_1)^2>0$), the phase difference ($\f$) and the propagation constant ($b$). The ratio $R$ is found by solving the following fourth-order polynomial equation:
\begin{eqnarray}
\left[\frac{\gamma}{\e}(\alpha R-1)+\beta_1(\beta-1)\frac{R}{R-1}\right]^2=\frac{k^2R}{4}-\alpha_1^2,
\label{QuarticEquation}
\end{eqnarray}
where $\alpha=-\alpha_2/\alpha_1>0$ and $\beta=\beta_2/\beta_1>0$. These ratios of the imaginary and real parts of the refractive indices describe the asymmetry of the structure and  are kept positive without loss of generality. Importantly, for $\alpha=\beta=1$ and $\e=0$, the system has a balanced gain and loss and the two waveguides have identical geometric properties determining their propagation constants thus corresponding to a Parity-Time (PT) symmetric configuration \cite{Ramezani_10}.  The ratio $R$ is restricted by the following conditions
 \begin{equation}
\left(\frac{k}{2\alpha_1}\right)^2 R>1
 \end{equation}
and
\begin{equation}
\gamma\beta_1(1-R)\left(\beta-1\pm\frac{\alpha_1}{\beta_1}\frac{1-R}{R}\sqrt{\left(\frac{k}{2\alpha_1}\right)^2 R-1}\right)>0
\end{equation}
reflecting the dependence of the asymmetry of the NS on the parameters of the structure and defining the regions of parameter space where NS exist. Moreover, it can be readily shown that $R=1/(\alpha-\e A_1^2)$. The other parameters characterizing the NS are given in terms of $R$ as follows
\begin{equation}
 \varphi =-\sin^{-1}\left( \frac{2 \alpha_1}{k \sqrt{R}}\right),
\end{equation}
\begin{equation}
 A_1^2 =\frac{1}{(1-R)\gamma}\left(\beta_1 (\beta-1)+\frac{k (1-R)}{2 \sqrt{R}}\cos{\varphi}\right),
\end{equation}
and
\begin{equation}
 b=\frac{1}{2}\left(\beta_1(\beta+1)+\gamma(R+1)A_1^2+\frac{k(R+1)}{2\sqrt{R}}\cos{\varphi}\right).
\end{equation}
In the absence of gain and loss saturation $\varepsilon=0$, the above expressions reduce to those obtained previously in \cite{Kominis_16, Kominis_17}.  It is worth emphasizing that there is a remarkable freedom in selecting the parameters of the system so that an asymmetric NS with arbitrary squared electric field amplitude ratio $R$ exists, allowing for directed power transport capabilities.  

The stability of the NS is determined by the eigenvalues of the Jacobian $\textbf{J}$ of the system evaluated at the specific NS 
\begin{equation}
\textbf{J}=\left[\begin{array}{ccc}
-\alpha_1                              & -\frac{k}{2}\sin\f                          & -\frac{k}{2}A_2\cos\f  \\
\frac{k}{2}\sin\f                      & -\alpha_2 \frac{1-\e A_2^2}{(1+\e A_2^2)^2} & \frac{k}{2}A_1\cos\f  \\
\frac{(R+1)k\cos\f}{2A_2}-2\gamma A_1 & 2\gamma A_2-\frac{(R+1)k\cos\f}{2A_1R}      & \frac{(R-1)k\sin\f}{2\sqrt{R}}
\end{array}\right],
\label{Jacobian}
\end{equation}
with a positive real part of at least one eigenvalue corresponding to instability of the respective NS.

\section{Results and Discussion}

\subsection{Parameter Space Analysis and Stability Maps}
Prior to the presentation of results on the existence and stability of NS, it would be meaningful to define value intervals for the quantities defining structure and excitation of the investigated coupled system. In the following, we restrict our analysis to the case of $\gamma>0$ corresponding to a self-focusing nonlinearity; the analysis is readily applicable to the case of self-defocusing nonlinearity $\gamma<0$ due to the invariance of the system (\ref{CME1})-(\ref{CME2}) under the ``staggering'' trasformation $\gamma \rightarrow -\gamma$, $\beta_{1,2} \rightarrow -\beta_{1,2}$, $E_1 \rightarrow -E_1^*$, $E_2 \rightarrow E_2^*$. Moreover, $\gamma$ is kept constant at $\gamma=1$. This choice is based on the fact that $A_1, A_2$ are normalized amplitudes usually possessing values of the order of unity; in this way, the nonlinear term gets comparable in magnitude with the linear one reinforcing the interplay between them. For similar reasons, we can use moderate values of the saturation parameter $\e$, in most cases within the interval: $0<\e<1$, to avoid purging the gain factor. The coupling coefficient $k$, which is common to both waveguides due to reciprocity, is inversely proportional to the distance between the two waveguides since it is achieved via the evanescent waves developed outside of them; thus, we assume $k/|\beta_1|\in[0.1, 10]$. Finally, the ratios $\alpha, \beta>0$, which indicate the asymmetry between the gain/loss distribution in the two waveguides and between the refractive indices of the two employed media respectively, cover quite an extensive range: $0.1<\alpha,\beta<10$. Here, we care more about the effect of the asymmetry on the two coupled waveguides rather than the quantitative propagation features into each of them individually; therefore, we assume unitary values for real and imaginary part of the wavenumber ($\alpha_1=\beta_1=1$) into the passive waveguide, without serious loss of generality.

\begin{figure}[ht!]
\centering
\subfigure[]{\includegraphics[width=4.1cm]{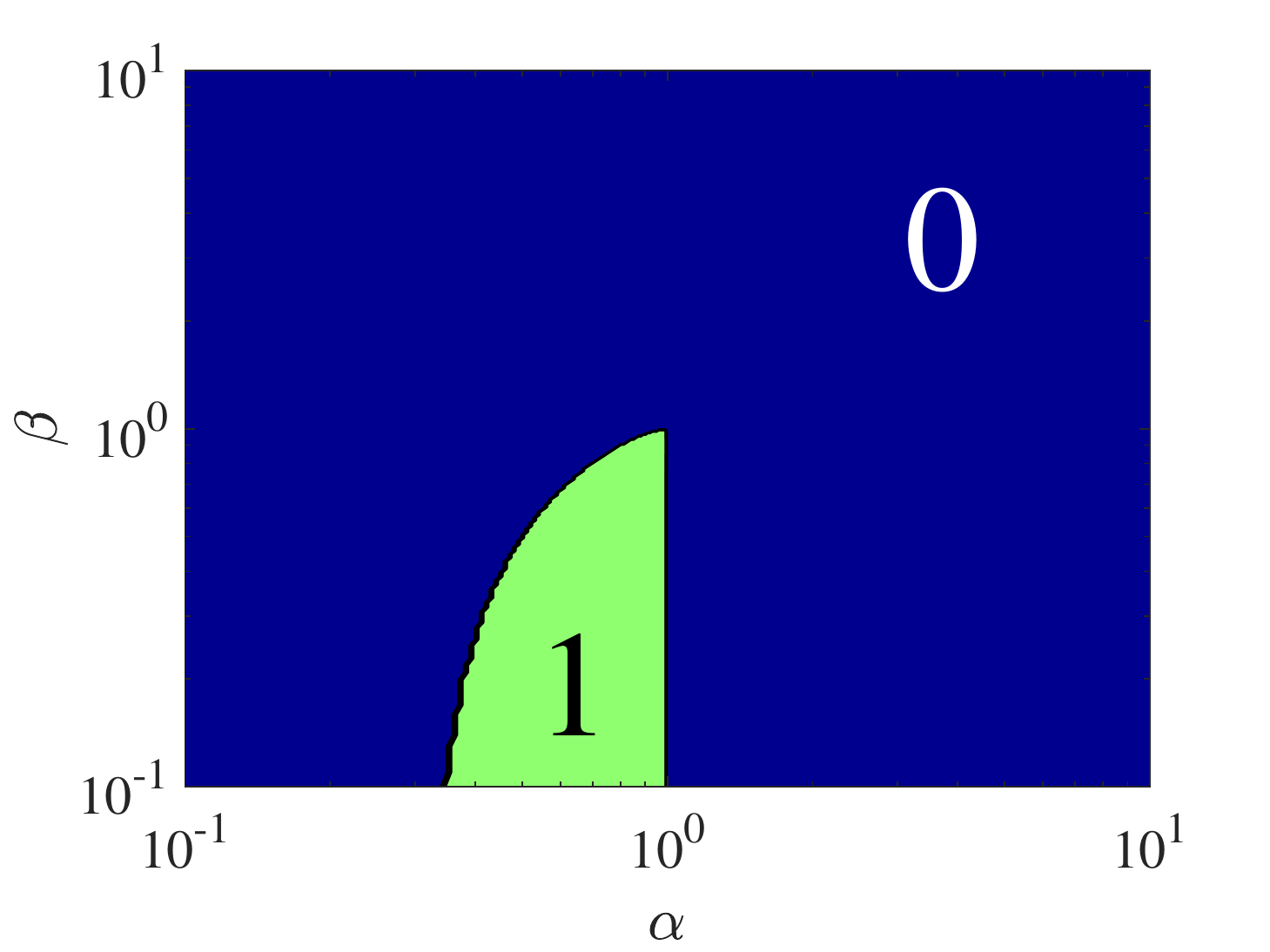}
   \label{fig:Fig1a}}
\subfigure[]{\includegraphics[width=4.1cm]{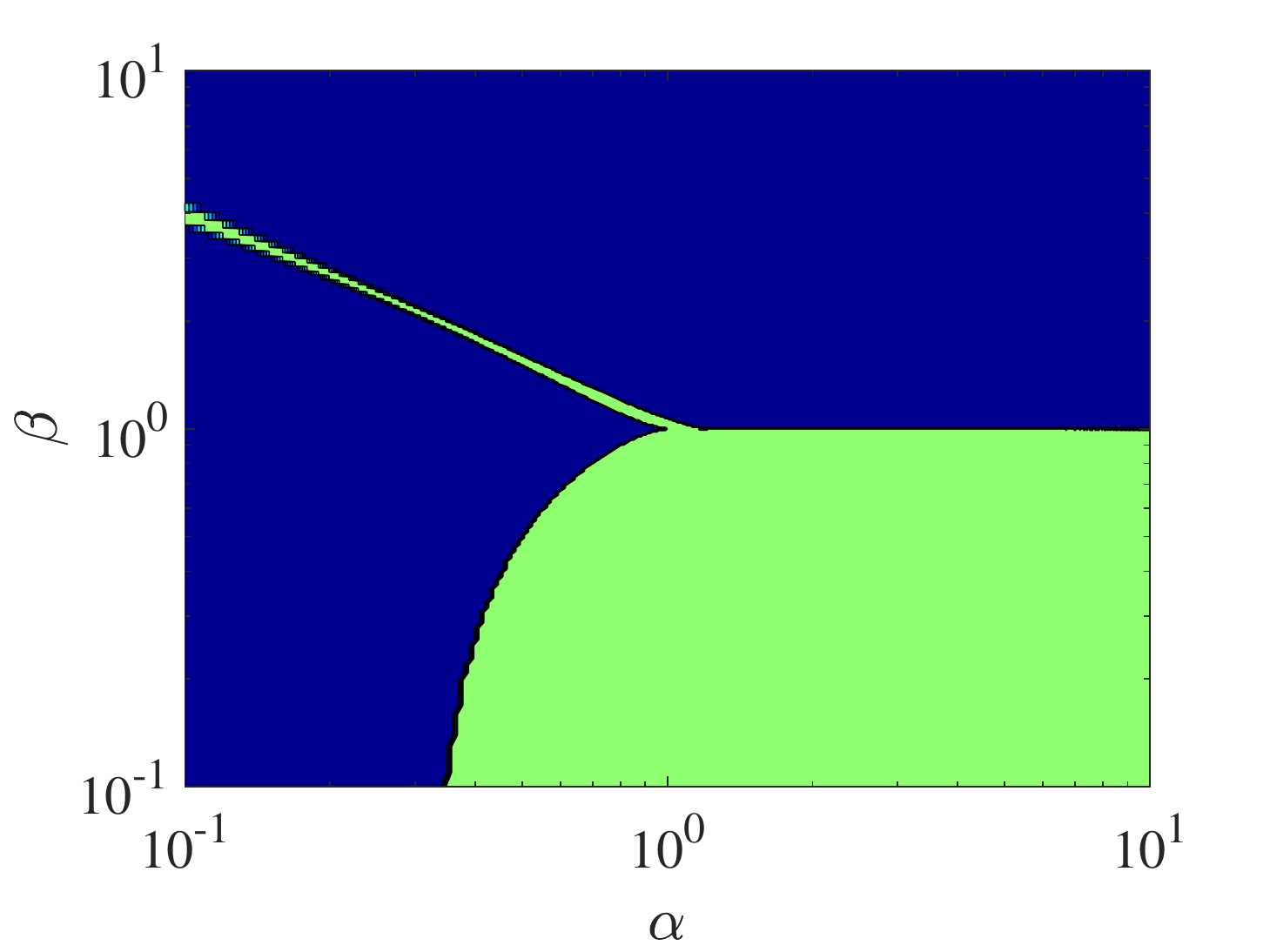}
   \label{fig:Fig1b}}\\
\subfigure[]{\includegraphics[width=4.1cm]{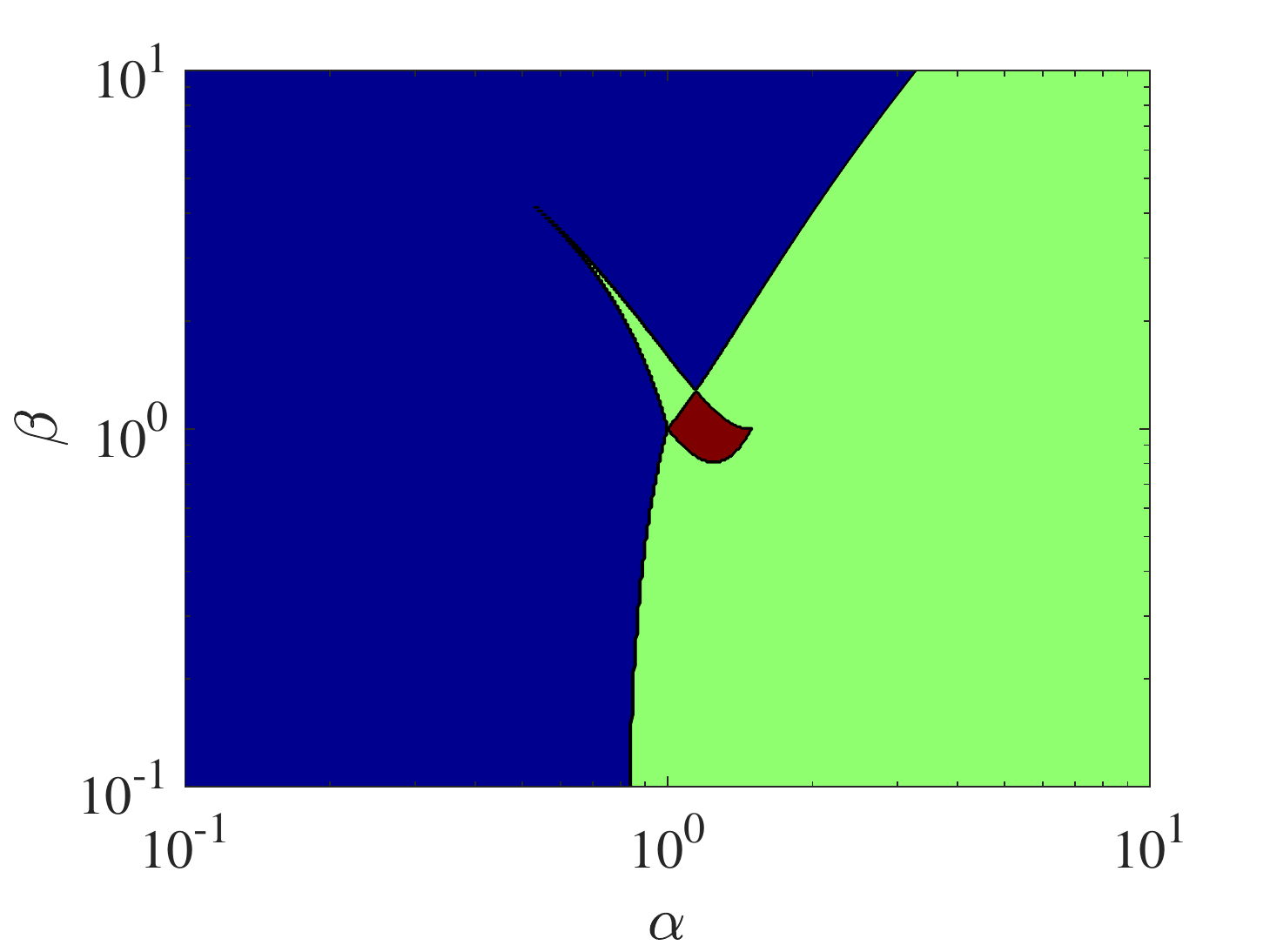}
   \label{fig:Fig1c}}
\subfigure[]{\includegraphics[width=4.1cm]{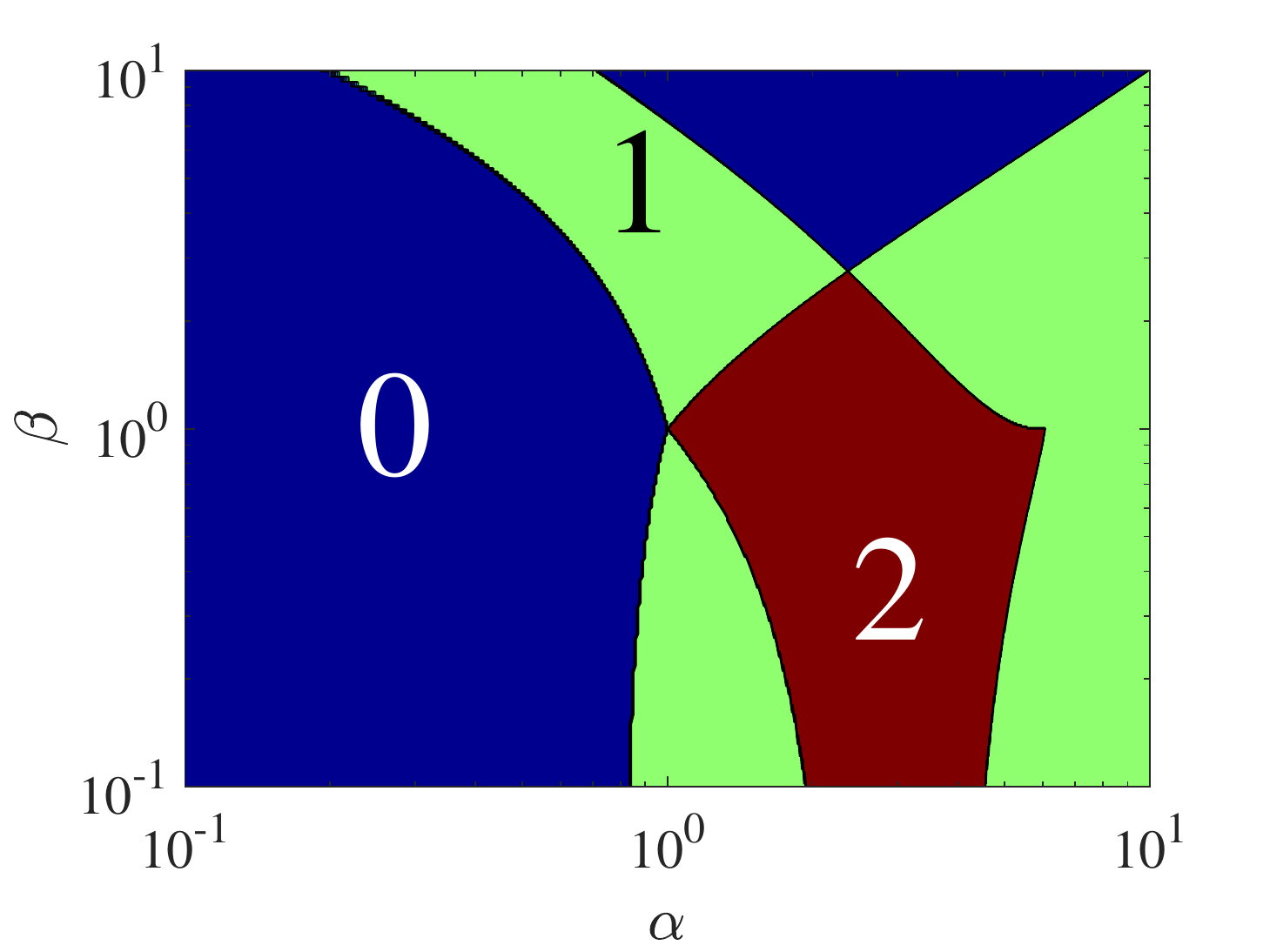}
   \label{fig:Fig1d}}
\caption{Number of stable NS (0-blue, 1-green, 2-brown) as function of gain/loss asymmetry $\alpha$ and propagation constants asymmetry $\beta$ for (a) zero saturation $\e=0$ and weak coupling $k/\beta_1=2$, (b) $\e=1$ and $k/\beta_1=2$, (c) $\e=0.1$ and $k/\beta_1=10$, (d) $\e=1$ and $k/\beta_1=10$.}
\label{fig:Figs1}
\end{figure}

The evolution of the system is crucially determined by the existence of stable NS and, in fact, any realistic application is directly related to the existence of at least one stable NS. However, apart from  converging to stable NS, the system may evolve either to a trivial zero mode  $(A_1=A_2=0)$ or to an unbounded state $(A_1=0, A_2\rightarrow+\infty)$ \cite{Ramezani_10, Kominis_16}.  In the following, we sweep the basic parameters and we obtain detailed stability maps depicting the number of stable NS in each region of parameter space. Regions with no stable NS are indicated by blue color, whereas regions with one or two stable NS are designated by green and red colors respectively.

\begin{figure}[ht!]
\centering
\subfigure[]{\includegraphics[width=4.1cm]{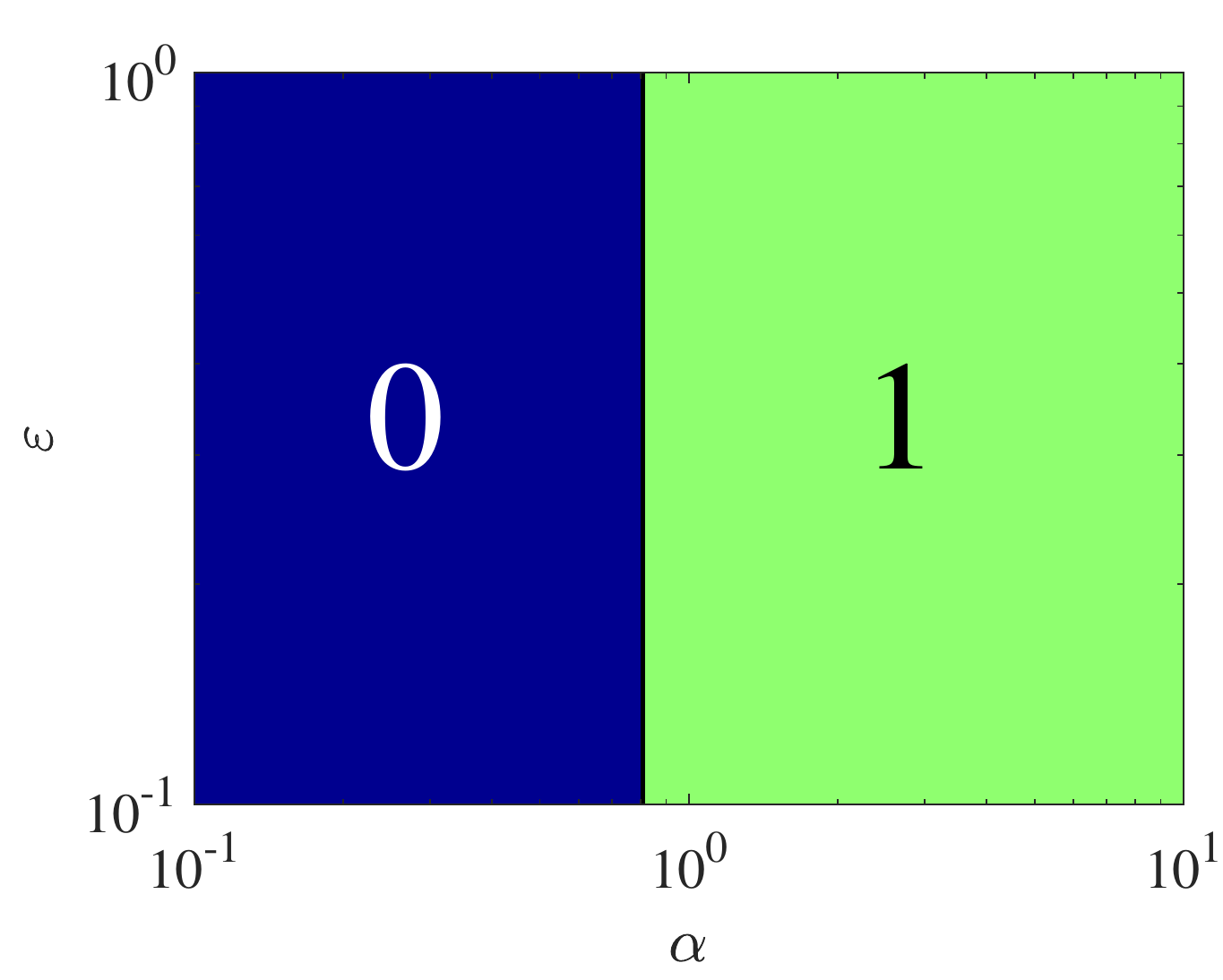}
   \label{fig:Fig2a}}
\subfigure[]{\includegraphics[width=4.1cm]{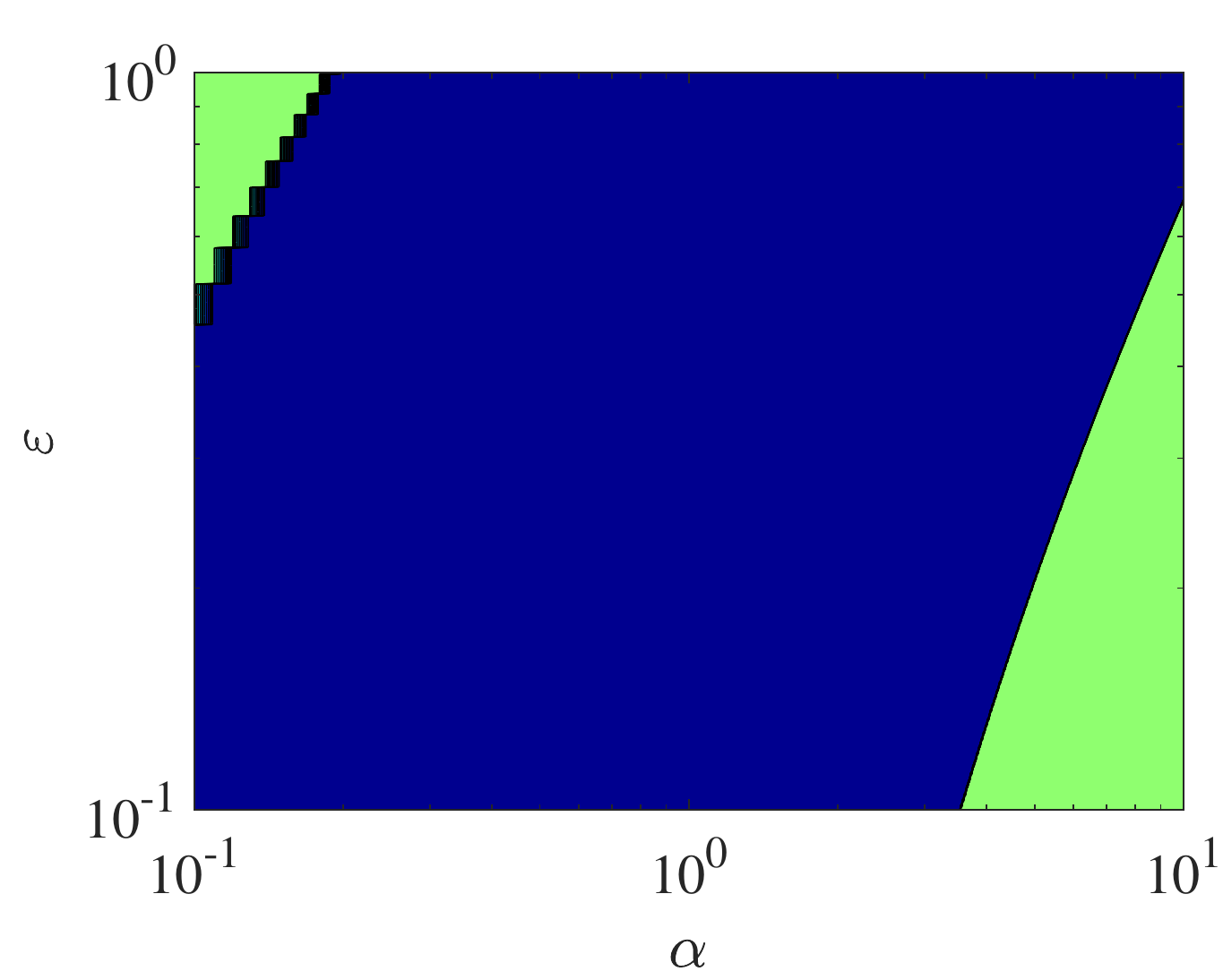}
   \label{fig:Fig2b}}\\
\subfigure[]{\includegraphics[width=4.1cm]{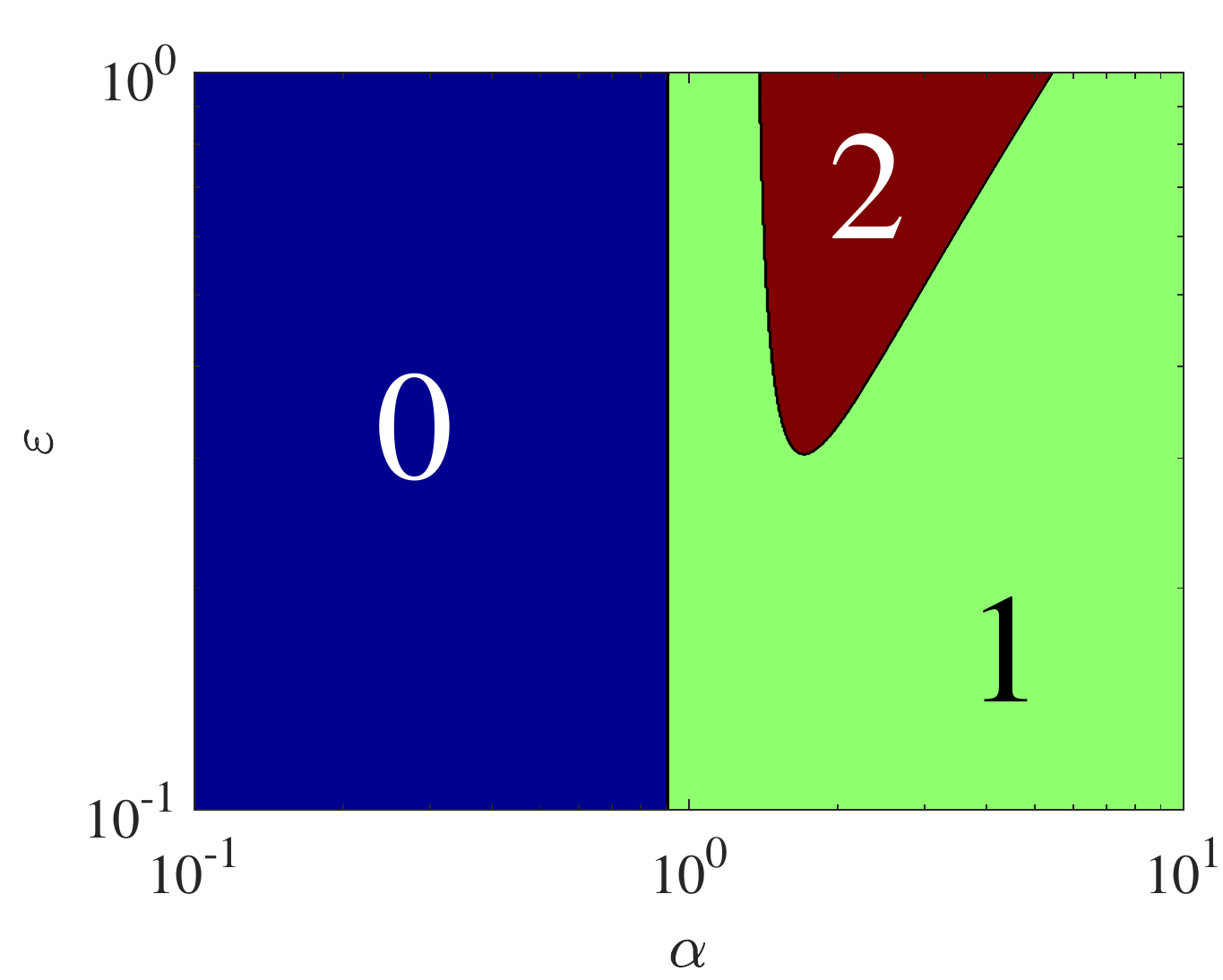}
   \label{fig:Fig2c}}
\subfigure[]{\includegraphics[width=4.1cm]{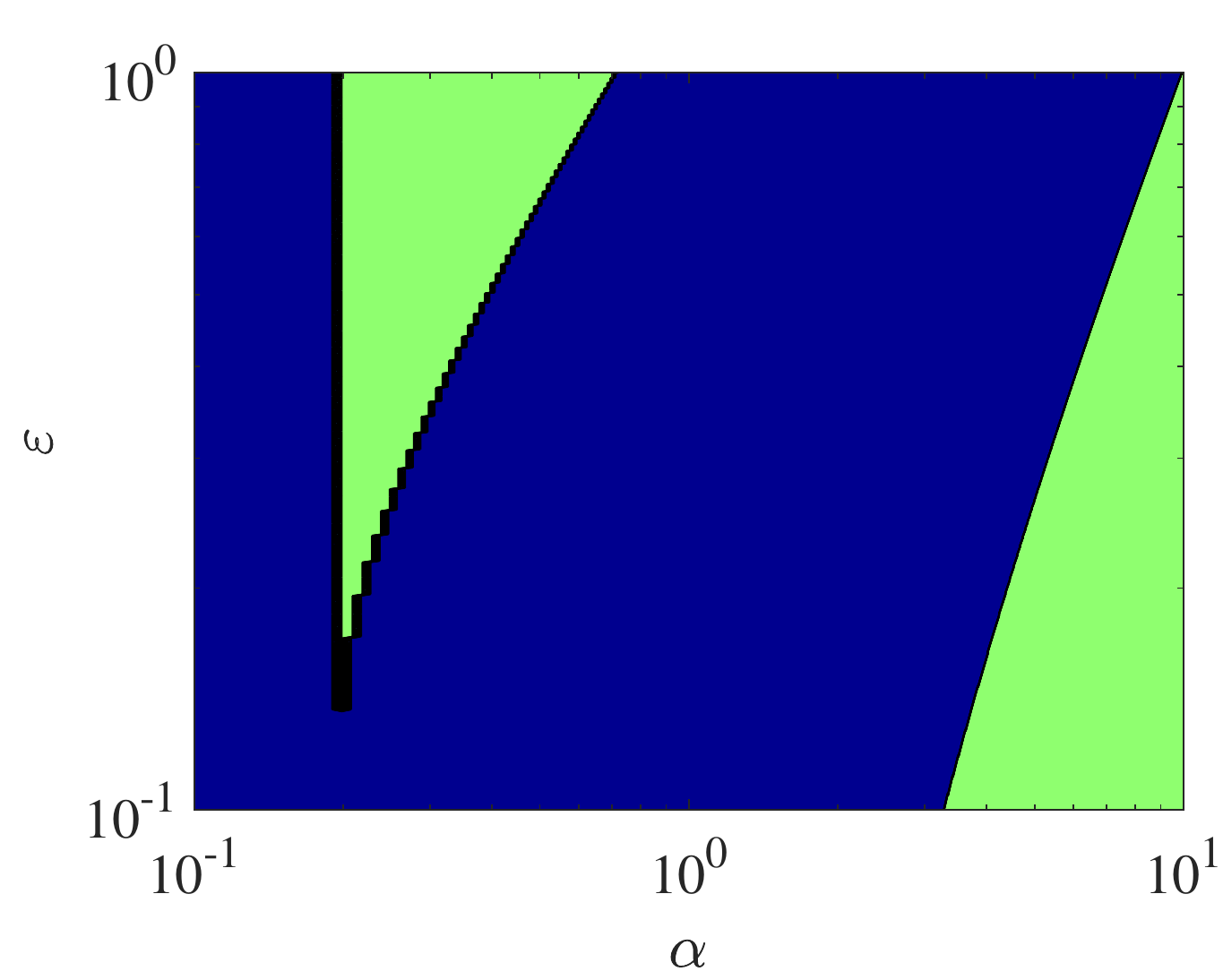}
   \label{fig:Fig2d}}
\caption{Number of stable NS (0-blue, 1-green, 2-brown) as function of gain/loss asymmetry $\alpha$ and saturation level $\e$ for (a) low propagation constant ratio $\beta=0.5$ and moderate coupling $k/\beta_1=5$, (b) $\beta=10$ and $k/\beta_1=5$, (c) $\beta=0.5$ and $k/\beta_1=10$, (d) $\beta=10$ and $k/\beta_1=10$.}
\label{fig:Figs2}
\end{figure}

In Figs \ref{fig:Figs1}, we show the number of stable NS supported by the system as a function of gain/loss contrast $\alpha$ and propagation constants asymmetry $\beta$ between the waveguides for various coupling and saturation levels. In Fig. \ref{fig:Fig1a}, we examine a system with no saturation ($\e=0$) and low coupling; it is noticed that stability is achieved only within a moderate zone of $\alpha$, while $\beta<1$. Remarkably, the behavior of the coupled waveguides ceases to be stable abruptly for $\alpha>1$, when the overall nature of the device becomes active. This is directly related to the symmetry of the dynamic equations \eqref{CME1}, \eqref{CME2} when $\e=0$, which becomes perfect at the PT-symmetric point ($\alpha=\beta=1$). One may wonder how is it possible to have a stable mode for larger values of gain $\alpha$ whereas no such mode exists for very small gain values. Such a trend is attributed to the strong engagement of the active waveguide with the passive one, due to their coupling interaction; for a very weak coupling, the active part is left alone to continuously increase its mode amplitude even for small gain values.

In Fig. \ref{fig:Fig1b}, we investigate the same setup as in Fig. \ref{fig:Fig1a} but with increased saturation ($\e=1$). One directly observes the beneficial influence of saturation on the stability response. In particular, the system can be stable for more substantial gain ranges $\alpha$; indeed, increased saturation ``clips'' the activity of the gain medium and prevents $A_2$ from divergence. Note that in Fig. \ref{fig:Fig1b} an additional and ultra-narrow stability region is opened for larger $\beta$, if properly matched with diminished gain/loss contrasts $\alpha$.  Importantly, the strong asymmetry between the two waveguides studied in Figs \ref{fig:Figs1}, is exploitable in directed power transport \cite{Kominis_16} and control of modulational instability \cite{Kominis_17}.


In Fig. \ref{fig:Fig1c}, we increase the coupling which enhances the collaboration towards stability of the two waveguides, compared with Figs \ref{fig:Fig1a} and  \ref{fig:Fig1b}. In addition, it leads to the onset of bistability across a small parametric region close to the PT-symmetric regime. Such a conclusion is further validated by Fig. \ref{fig:Fig1d}, where the same sizeable coupling ($k/\beta_1=10$) is combined with significant saturation ($\e=10$). A much more extended parametric ``plateau'' of bistability is formulated, accompanied by substantial restriction of unstable regions. By inspection of Figs \ref{fig:Fig1c} and \ref{fig:Fig1d}, we remark that bistability domains never share a common boundary with domains where no stable NS exist; they only touch each other via isolated points. The bistability reported in Figs \ref{fig:Fig1c} and \ref{fig:Fig1d}, accompanied by hysteresis, is indispensable for information processing in various setups including acoustic \cite{CascadingElasticVibrations} or thermal \cite{ThermalMemory} components, and for photonic memory primarily in optical structures \cite{FilmCoupledMetasurfaces}. 

\begin{figure}[ht!]
\centering
\subfigure[]{\includegraphics[width=4.1cm]{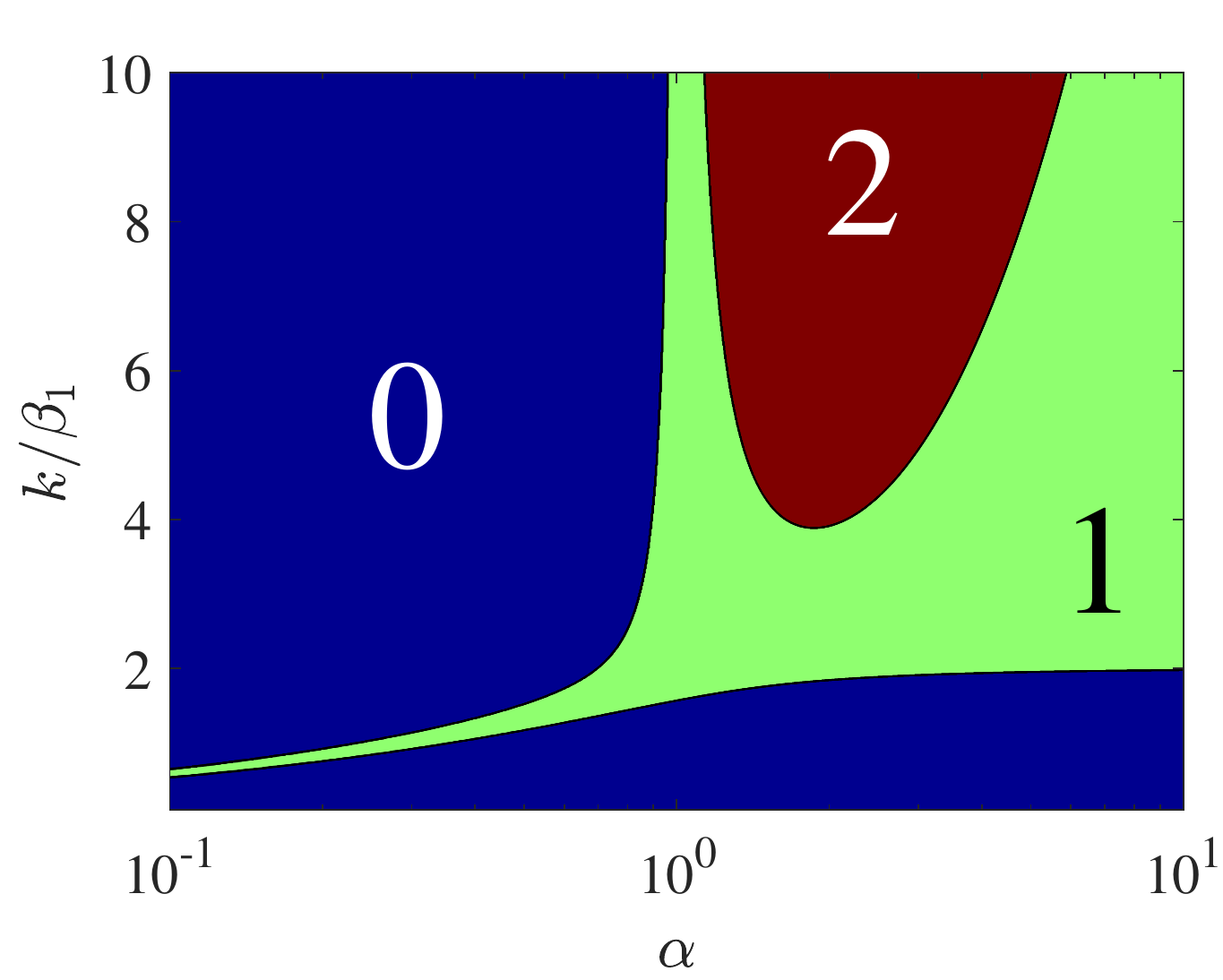}
   \label{fig:Fig23a}}
\subfigure[]{\includegraphics[width=4.1cm]{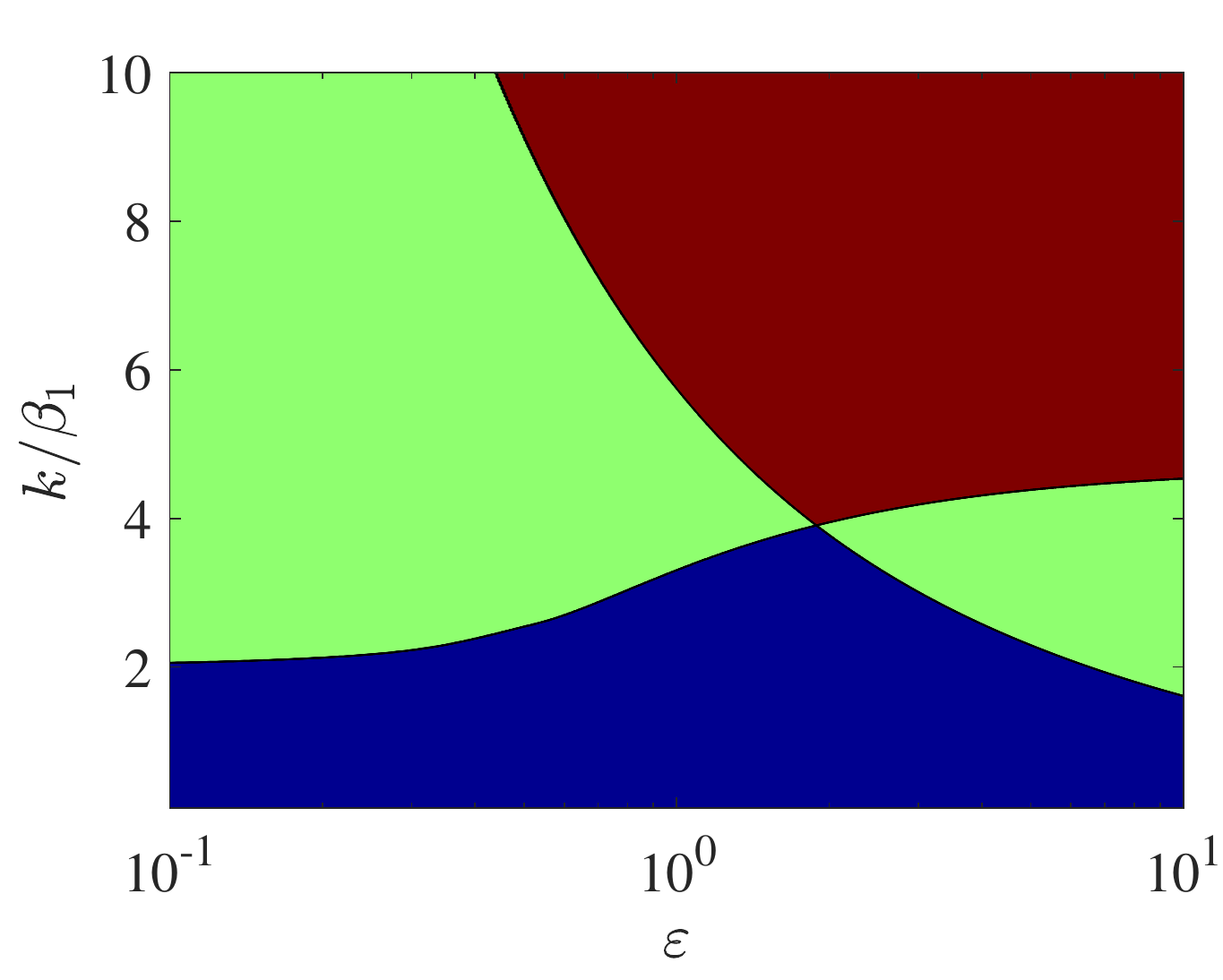}
   \label{fig:Fig23b}}
\caption{Number of stable NS (0-blue, 1-green, 2-brown) as a function of normalized coupling $k/\beta_1$ and: (a) gain/loss asymmetry $\alpha$ (with $\beta=0.8$, $\e=1$), (b) saturation coefficient $\e$ (with $\beta=1.5$, $\alpha=2$).}
\label{fig:Figs23}
\end{figure}

In Figs \ref{fig:Figs2} we display stability maps, as in Figs \ref{fig:Figs1}, where we scan the whole range of saturation coefficient  $\e$. In Fig. \ref{fig:Fig2a}, where low values of $\beta$ and $k$ are assumed, we observe an abrupt change in the stability behavior occurring at a specific value of $\alpha\cong 0.91$. In Fig. \ref{fig:Fig2b}, we further increase $\beta$ and note that stability is also achievable for low levels of $\alpha$, unlike in Fig. \ref{fig:Fig2a}; however, the system becomes overall more unstable (bigger blue region compared to Fig. \ref{fig:Fig2a}). In Fig. \ref{fig:Fig2c}, where the coupling is stronger, bistability emerges for a considerable part of the stable parametric space, even though the unstable region remains almost unchanged. In Fig. \ref{fig:Fig2d} we examine the same pair of waveguides as in Fig. \ref{fig:Fig2b} but at smaller distance between them (larger $k/\beta_1=10$); such a design modification allows the system to support a stable regime for very low saturation levels even when $\alpha$ is below unity. Stability of saturable couplers can be efficiently utilized as cells in computing photonic platform \cite{InMemory} or in examining atom-photon coupling with high radiative decay \cite{AtomLightInteractions}.

In Figs \ref{fig:Figs23} we investigate how sweeping the range of $k$ values affects the stability of the system and thus represent the number of stable NS on maps whose vertical axis indicates the normalized variable $k/\beta_1$ (with $\beta_1=1$). In Fig. \ref{fig:Fig23a}, we observe that, for small gain $\alpha$, the existence of a stable NS is restricted in a thin parametric strip of the $(\alpha,k)$ map. However, if gain/loss asymmetry $\alpha$ (for $\alpha_1=1$) exceeds a threshold, the coupling favors the convergence of solutions and that is why large stability and bistability parametric domains emerge. In Fig. \ref{fig:Fig23b}, where the range of saturation coefficient $\e$ has been extended, we notice the absence of stable NS for $k\rightarrow 0$.  Furthermore, we again verify the findings shown in Figs \ref{fig:Fig1c} and \ref{fig:Fig1d}; indeed, the transition from two stable NS to none happens only through isolated points, unlike the transition from regions with 2 stable NS to 1 or from 1 to 0, which have common borders. Finally, Fig. \ref{fig:Fig23b} illustrates the crucial role of saturation on the bistability of the coupler; importantly, it seems that bistability is not feasible in the absence of saturation ($\e=0$). 

\begin{figure}[ht!]
\centering
\subfigure[]{\includegraphics[width=7.1cm]{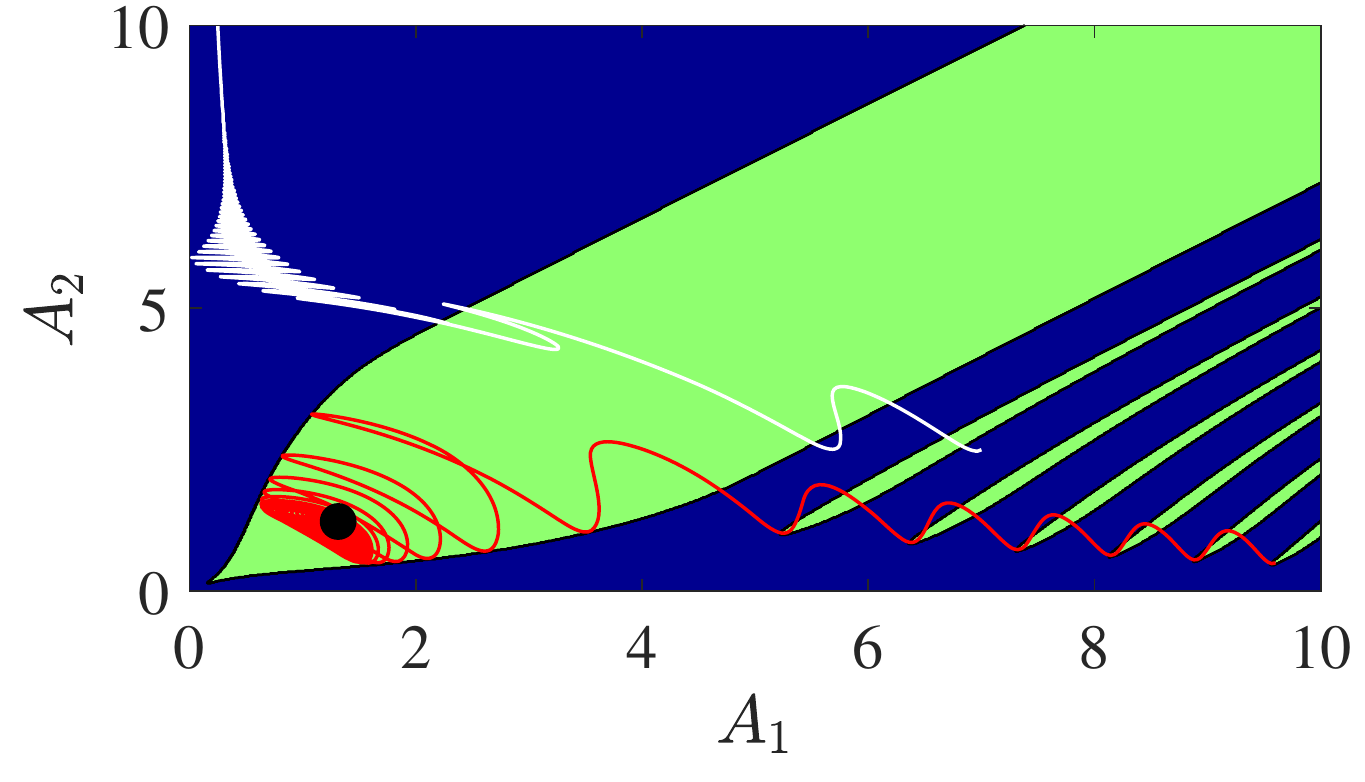}
   \label{fig:Fig3a}}\\
\subfigure[]{\includegraphics[width=4.1cm]{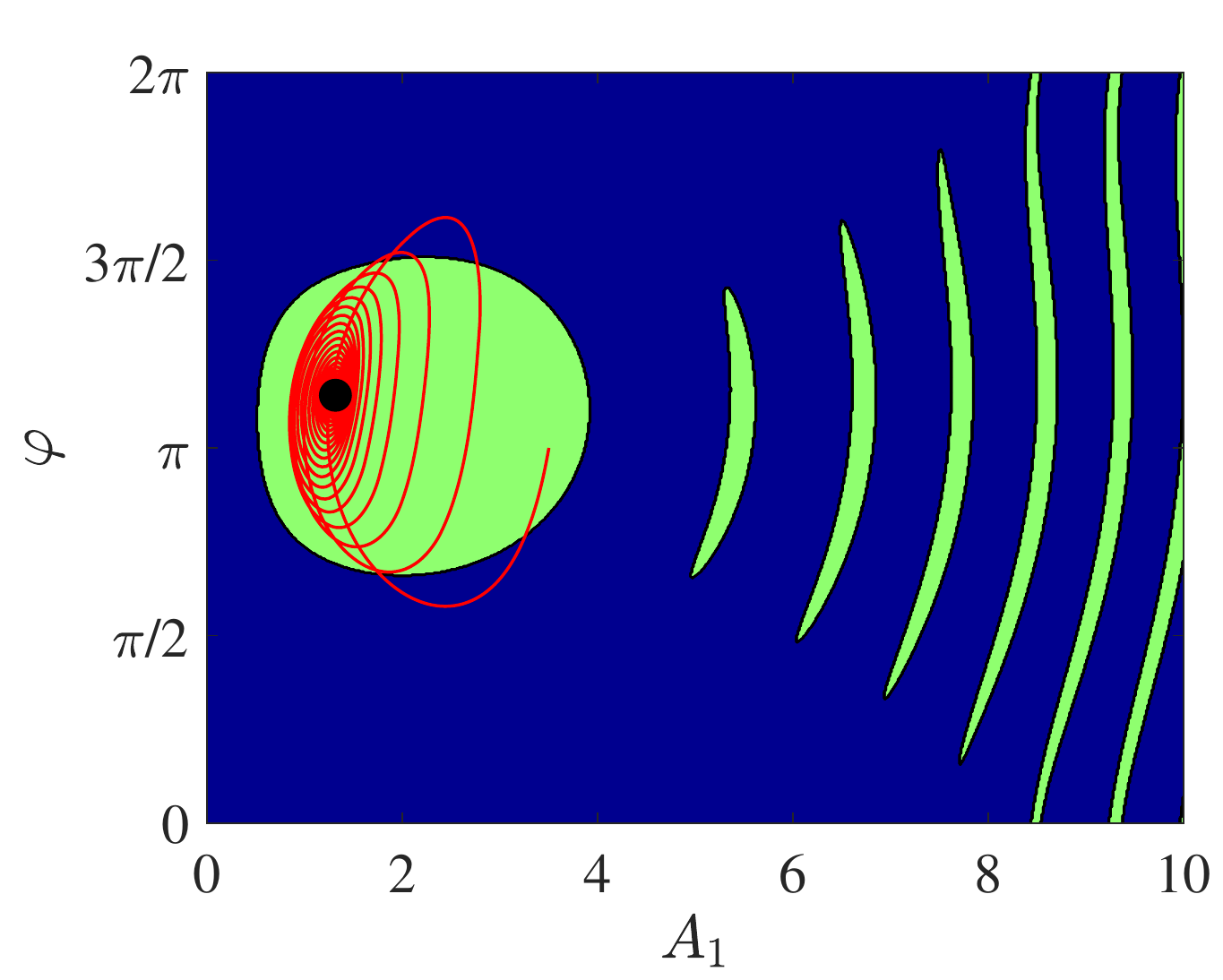}
   \label{fig:Fig3b}}
\subfigure[]{\includegraphics[width=4.1cm]{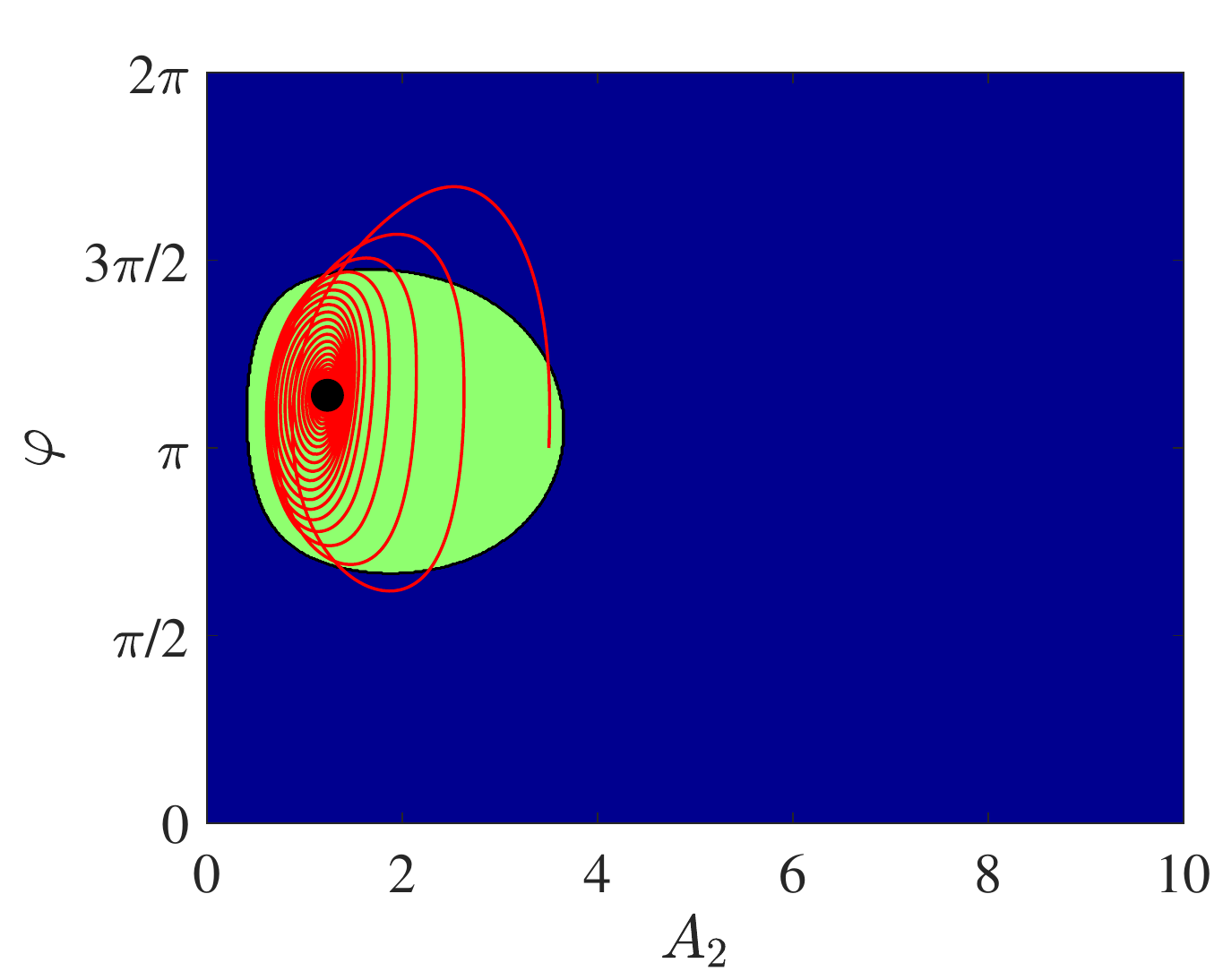}
   \label{fig:Fig3c}}
\caption{Basins of attraction and representative solution trajectories on the initial condition planes. (a) $(A_1(0),A_2(0))$ map for $\f(0)=\bar{\f}$, (b) $(A_1(0),\f(0))$ map for $A_2(0)=\bar{A}_2$, (c) $(A_2(0),\f(0))$ map for $A_1(0)=\bar{A}_1$. The basin of attraction of the stable NS and the instability region are marked by green and blue color respectively. Plot parameters: $\alpha=2$, $\beta=1.5$, $\e=0.5$  and $k/\beta_1=5$.}
\label{fig:Figs3}
\end{figure}

\subsection{Phase Space Analysis and Basins of Attraction}
The existence of a stable NS does not guarantee the evolution of the system to this state; the system may evolve either to the stable zero state, an unbounded state or a coexisting pair of stable NS in the case of bistability, depending on the initial conditions $\{A_1(0),A_2(0),\f(0)\}$. By selecting the parameters of the system so that at least one stable NS exists we exclude the possibility of the system's evolution to the zero state, since it has been found to be stable only at parameter regions where no stable NS exists \cite{Kominis_16}. However, evolution to a specific stable NS (or the undesirable unbounded state) still depends strongly on the initial conditions. Each stable NS is associated with a basin of attraction, defined as the set of initial conditions that asymptotically converge to this point. The extent of the basin of attraction of each stable NS in the system's phase space provides a measure of  how ``attractive'' this NS is. In what follows, we show characteristic cases of basins of attraction calculated numerically by considering a fine grid of initial conditions on specific plane cuts of the three-dimensional phase space and characterize them according to their asymptotic evolution.

Such basins of attraction are sketched in Fig. \ref{fig:Fig3a} for the case  of a single stable NS $\{\bar{A}_1,\bar{A}_2,\bar{\f}\}$ on an $(A_1,A_2)$ map; the initial value $A_1(0)$ is represented along the horizontal axis and $A_2(0)$ along the vertical one. The third quantity, the phase difference, is kept fixed and equal to the steady-state value, namely $\f(0)=\bar{\f}$. Green color marks the basin of attraction of the stable NS which is indicated by a black dot. Similarly, blue color marks the basin of attraction of the unbounded state (instability region). It is noteworthy that apart from the main area containing the stable NS, the basin of attraction has a fine structure consisting of several thin strips, for larger $A_1(0)$, demonstrating the sensitive dependence on initial conditions, typical in complex nonlinear systems. To demonstrate the utility of the presented information, we show a couple of indicative trajectories of the system corresponding to initial conditions evolving asymptotically to the stable NS and the unbounded state. The red line corresponds to the solution trajectory beginning from one of the thin stable zones far away from the fixed point. We notice that, after executing some oscillations and passing through instability domains (at which apparently $\f(z)\ne\bar{\f}$), the system converges to the steady-state solution. The case of initial conditions chosen within the blue region is illustrated by a white line in Fig. \ref{fig:Fig3a}. Indeed, after some spatially abrupt changes, $A_2$ and $A_1$ tend monotonically to infinity and zero values respectively.

In Fig. \ref{fig:Fig3b} we show the basins of attraction of the same system as in Fig. \ref{fig:Fig3a}, but across a different plane cut: the phase difference $\f(0)$ at the vertical axis. It should be stressed that the two sketches (Figs \ref{fig:Fig3a} and \ref{fig:Fig3b}) do not contain redundant information since this time the amplitude $A_2$ has been pre-selected equal to its convergent value, namely $A_2(0)=\bar{A}_2$. We display a stable trajectory (red line) which, after following a spiral route, converges to the fixed point (black dot). In Fig. \ref{fig:Fig3c} we draw again the basins of attraction for the same configuration on $(A_2,\f)$ plane by choosing the initial amplitude $A_1$ equal to $\bar{A}_1$. We observe that in this plane cut the basin of attraction has a simply connected topology.


\begin{figure}[ht!]
\centering
\subfigure[]{\includegraphics[width=7.1cm]{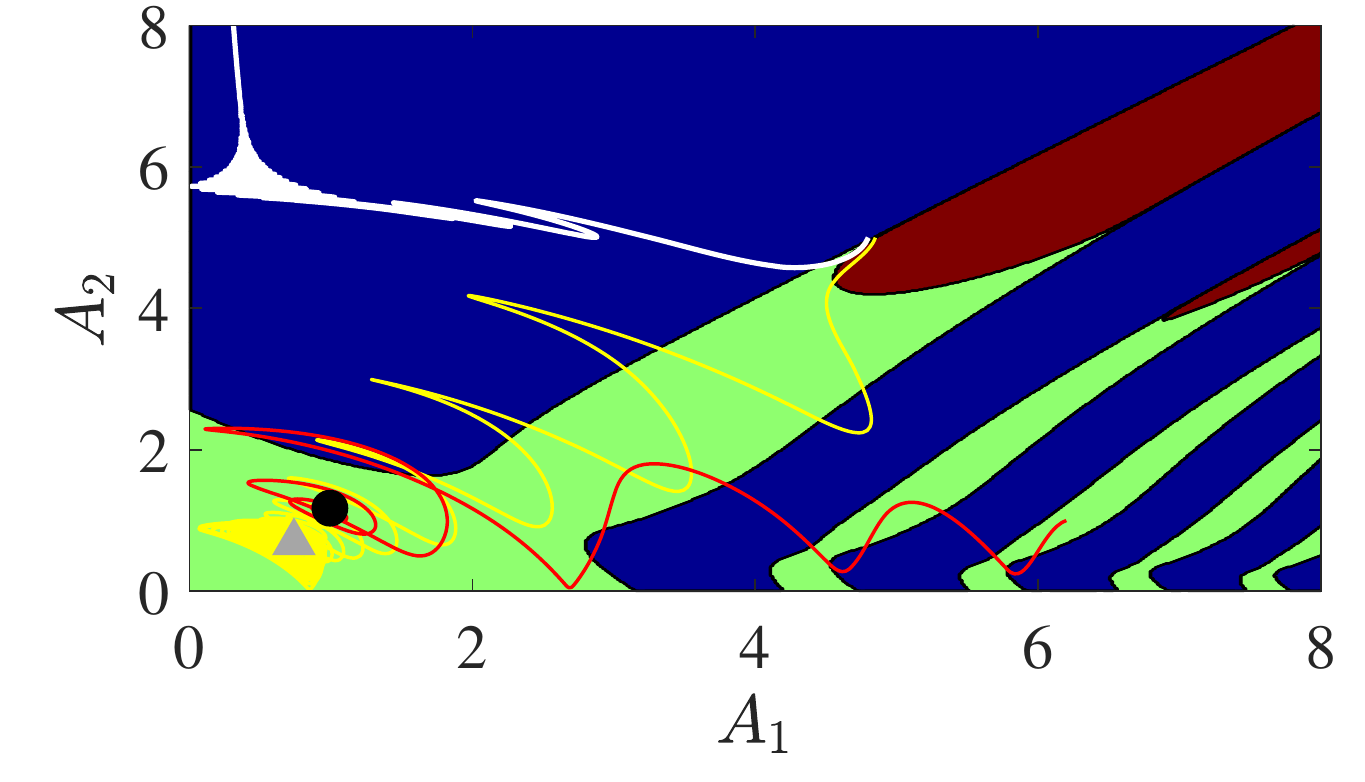}
   \label{fig:Fig4a}}\\
\subfigure[]{\includegraphics[width=4.1cm]{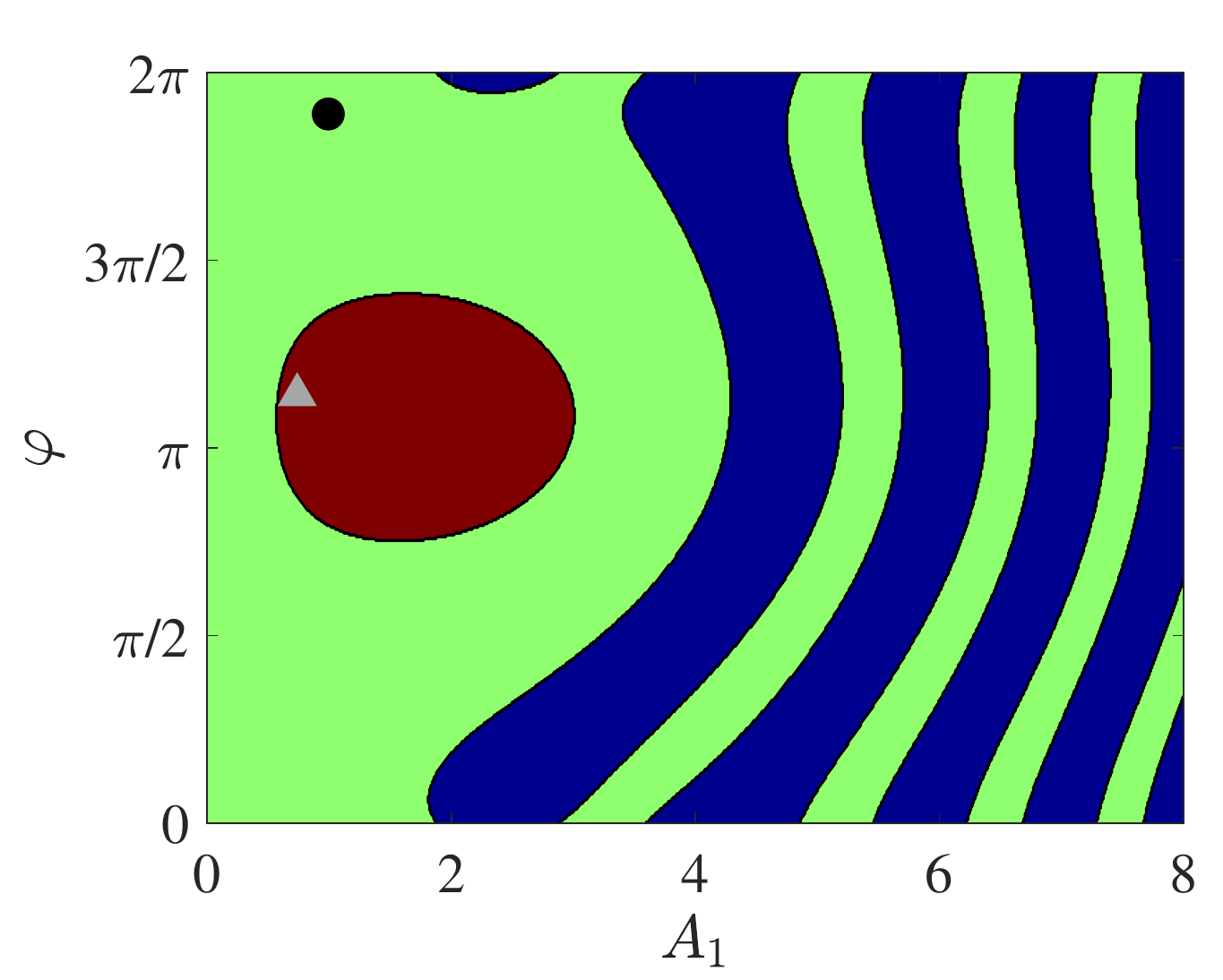}
   \label{fig:Fig4b}}
\subfigure[]{\includegraphics[width=4.1cm]{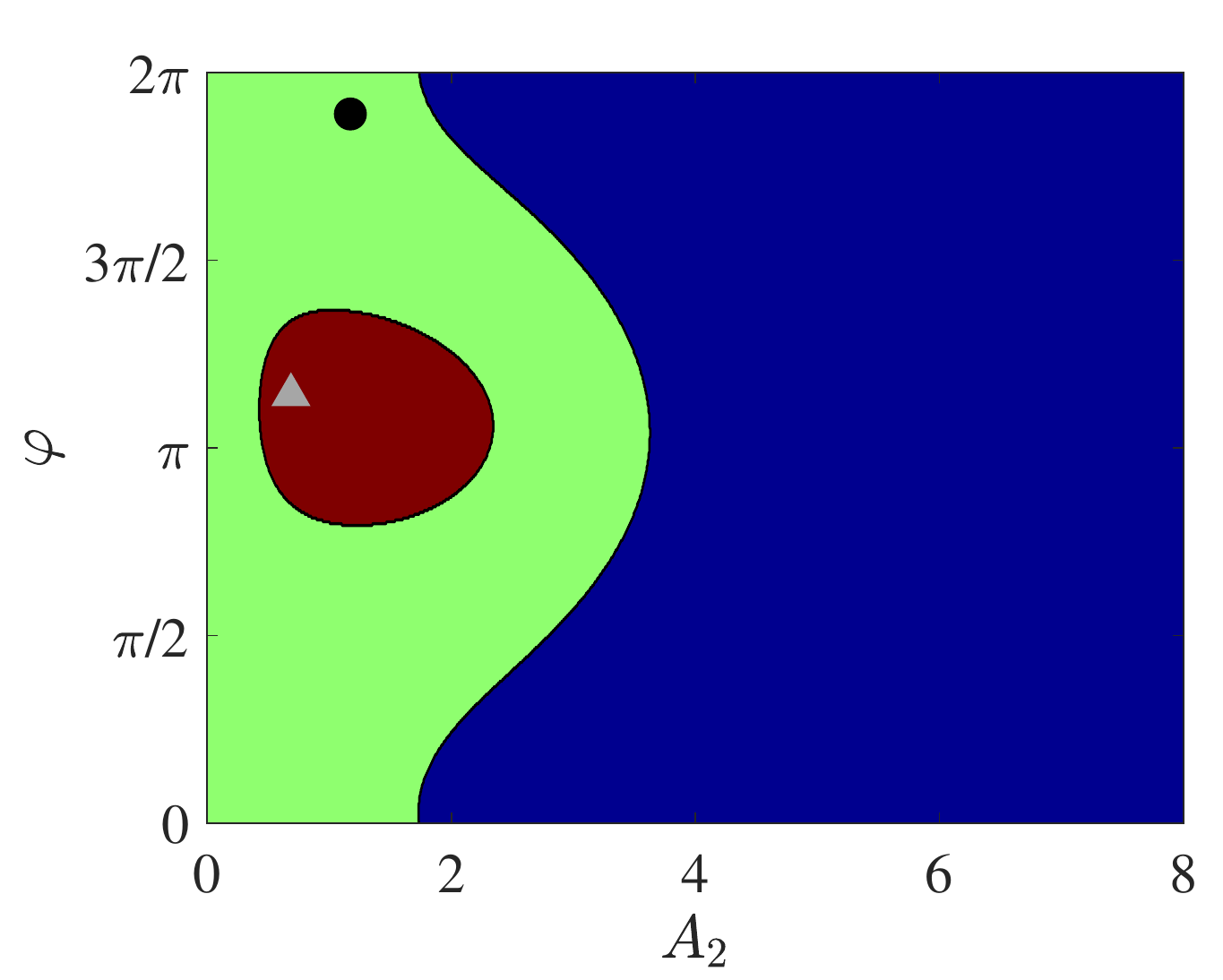}
   \label{fig:Fig4c}}
\caption{Basins of attraction and representative solution trajectories on the initial condition planes of a system with two stable fixed points for: (a) $\f(0)=\bar{\f}$, (b) $A_2(0)=\bar{A}_2$, (c) $A_1(0)=\bar{A}_1$. The basins of attraction of the two stable NS and the instability region are marked by green, brown and blue color respectively. Plot parameters: $\alpha=1.7$, $\beta=1.4$, $\e=1$ and $k/\beta_1=5$.}
\label{fig:Figs4}
\end{figure}

In Figs \ref{fig:Figs4}, we investigate a coupled layout supporting two stable points denoted as: $\{\bar{A}_1,\bar{A}_2,\bar{\f}\}$ and $\{\hat{A}_1,\hat{A}_2,\hat{\f}\}$. Therefore, each triplet of initial conditions $\{A_1(0),A_2(0),\f(0)\}$ may give rise to a solution converging to one (green color) or the other (red color) stable NS, or lead to instability. As implied above, the basins of attraction sketched on a two-dimensional plane cut require an implicit selection for the third parameter of our three-dimensional phase space, which has been taken equal to the corresponding value of the fixed point (like $\f(0)=\bar{\f}$ in Fig. \ref{fig:Fig3a}). When bistability occurs, there are two alternatives for this selection (first or second stable NS) leading to different forms of basins of attraction shown in each plane cut.

In Fig. \ref{fig:Fig4a}, we represent two fixed points: $\{\bar{A}_1,\bar{A}_2,\bar{\f}\}$ (black dot) and $\{\hat{A}_1,\hat{A}_2,\hat{\f}\}$ (gray triangle) on $(A_1,A_2)$ plane, implying that $\f$ is respectively different; indeed, the two stable states are not as close to each other as in Fig. \ref{fig:Fig4a}, since $\bar{\f}\ne \hat{\f}$. However, the domains of stability/instability are computed considering that the initial phase difference $\f$ is kept equal to that of the first NS: $\f(0)=\bar{\f}$. A major locus of stability is established, which is divided between the two stable NS; it is again accompanied by thin strips similar to those of Fig. \ref{fig:Fig3a}, but bi-colored this time. 

In Fig. \ref{fig:Fig4a} we also show characteristic evolutions of $A_1(z),A_2(z)$; one can observe a case converging to the first NS $\{\bar{A}_1,\bar{A}_2,\bar{\f}\}$ (red line), in a similar manner as in Fig. \ref{fig:Fig3a}. In addition, we examine two neighboring initial points on a $(A_1,A_2)$ plane exhibiting different behaviors for $z\rightarrow+\infty$ (one unstable, another converging to the second NS). It is striking how they diverge from each other, with the first one (white line) following the way towards instability and the second (yellow line) spiraling around $\{\hat{A}_1,\hat{A}_2,\hat{\f}\}$.

In Figs \ref{fig:Fig4b} and \ref{fig:Fig4c}, we calculate the basins of attraction with $A_2(0)=\bar{A}_2$ and $A_1(0)=\bar{A}_1$ respectively, as in Figs \ref{fig:Fig3b} and \ref{fig:Fig3c}. Obviously, the patterns are also $2\pi$-periodic with respect to $\f$. Again, elongated segments of convergence to the first NS appear alternating with instability stripes. Furthermore, one can achieve convergence to $\{\bar{A}_1,\bar{A}_2,\bar{\f}\}$ even for $A_1(0)\rightarrow 0$, without caring much about $\f(0)$ as long as $A_2(0)$ is properly selected; on the contrary, one can reach $\{\hat{A}_1,\hat{A}_2,\hat{\f}\}$ only for a specific range of $\f(0)$, a result attributed to the choice $A_1(0)=\bar{A}_1$.

\begin{figure}[ht!]
\centering
\subfigure[]{\includegraphics[width=6.4cm]{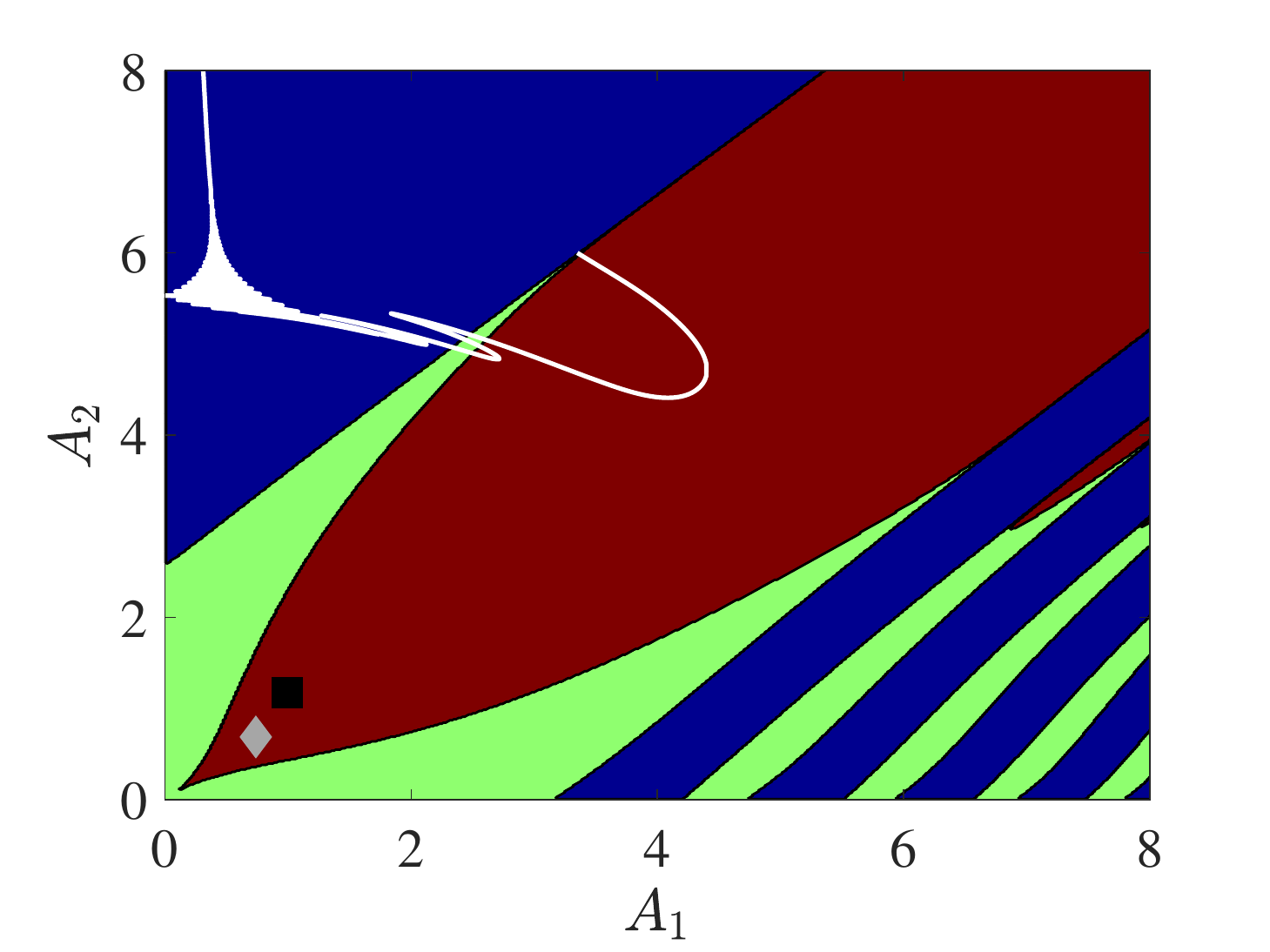}
   \label{fig:Fig5a}}
\subfigure[]{\includegraphics[width=6.4cm]{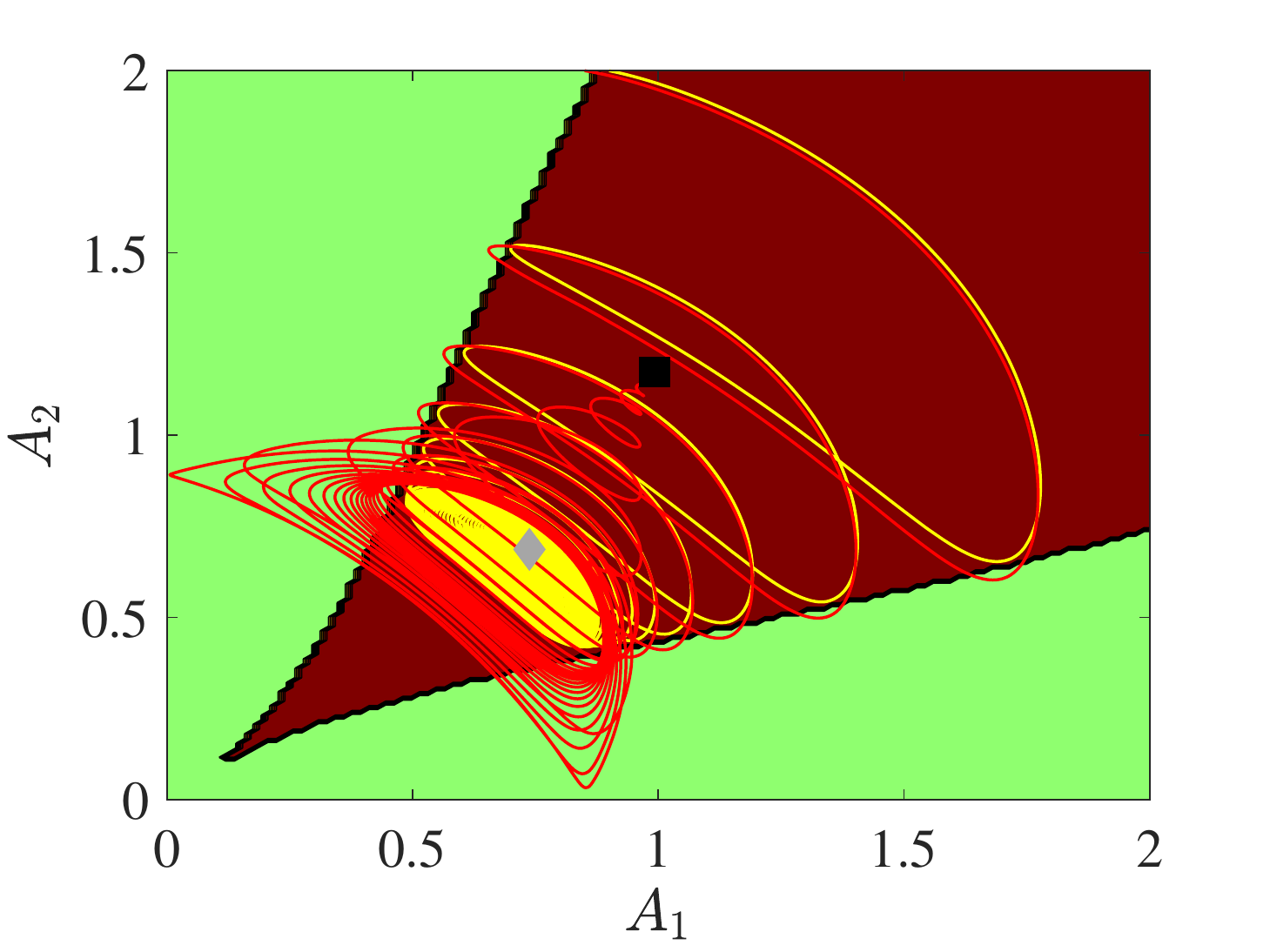}
   \label{fig:Fig5b}}
\caption{Basins for attraction of the same system as in Figs \ref{fig:Figs4} but with initial phase difference equal to the steady-state value of the other fixed point, $\f(0)=\hat{\f}$. (a) $(A_1(0),A_2(0))$ map, (b) detail of $(A_1(0),A_2(0))$ map around the two fixed points. The basins of attraction of the two stable NS and the instability region are marked by green, brown and blue color respectively. }
\label{fig:Figs5}
\end{figure}

In Figs \ref{fig:Figs5} we sketch the basins of attraction for the same system of Figs \ref{fig:Figs4} but with a different implicit selection of the third initial condition. In particular, in Fig. \ref{fig:Fig5a}, we show the stability regions on $(A_1(0),A_2(0))$ map with $\f(0)=\hat{\f}$; because of this alternative choice of $\f(0)$ compared to Fig. \ref{fig:Fig4a}, the final states of the system are designated by the black square and gray rhombus. We emphasize that the shape of the domains is totally different, a feature which is again accredited to the three-dimensional nature of the problem making the two-dimensional cross cuts of convergence volumes for fixed $\f(0)$ (or one of the other two quantities $A_1(0)$, $A_2(0)$), to be dependent on $\f(0)$ (or $A_1(0)$, $A_2(0)$ respectively). In particular, the extent of the basin of attraction corresponding to the first stable NS shrinks, whereas the other one dominates in contrast to Fig. \ref{fig:Fig4a}. Again, we show a characteristic trajectory for an initial condition chosen very close but outside the basin of attraction of the stable NS; as in the corresponding trajectories of Figs \ref{fig:Fig3a}, \ref{fig:Fig4a}, the system evolves to the unbounded state with $A_1(z)\rightarrow 0$ and  $A_2(z)\rightarrow+\infty$.

In Fig. \ref{fig:Fig5b}, we show a detail of the bottom left corner from the map of Fig \ref{fig:Fig5a}, together with two additional trajectories starting from two almost adjacent initial points but belonging to different basins of attraction: one to the first NS (green region, red line) and the other to the second NS (red region, yellow line). The projections of the two trajectories almost coincide initially; as $z$ increases, the one starting from the basin of attraction of the second NS converges fast to the corresponding steady state with its amplitude undergoing multiple oscillations. However, the trajectory in red color sticks on the second NS (gray rhombus) winding around it many times before finally reaching its own steady state (black square). The characteristic of a solution spending most part of its trajectory around the other steady state $\{\hat{A}_1,\hat{A}_2,\hat{\f}\}$ instead of its own $\{\bar{A}_1,\bar{A}_2,\bar{\f}\}$ seems definitely correlated with our initial phase difference choice equal $\f(0)=\hat{\f}$. Such behavior of trajectories ``sticking'' at a point different from that of the final equilibrium is typical for complex nonlinear systems \cite{Thompson}. It is finally clear that small perturbations in initial conditions can cause the coupler response to move between the coexisting attractors, resulting in wild fluctuations in the power output \cite{ComplicatedBasins}.

\begin{figure}[ht!]
\centering
\subfigure[]{\includegraphics[width=4.1cm]{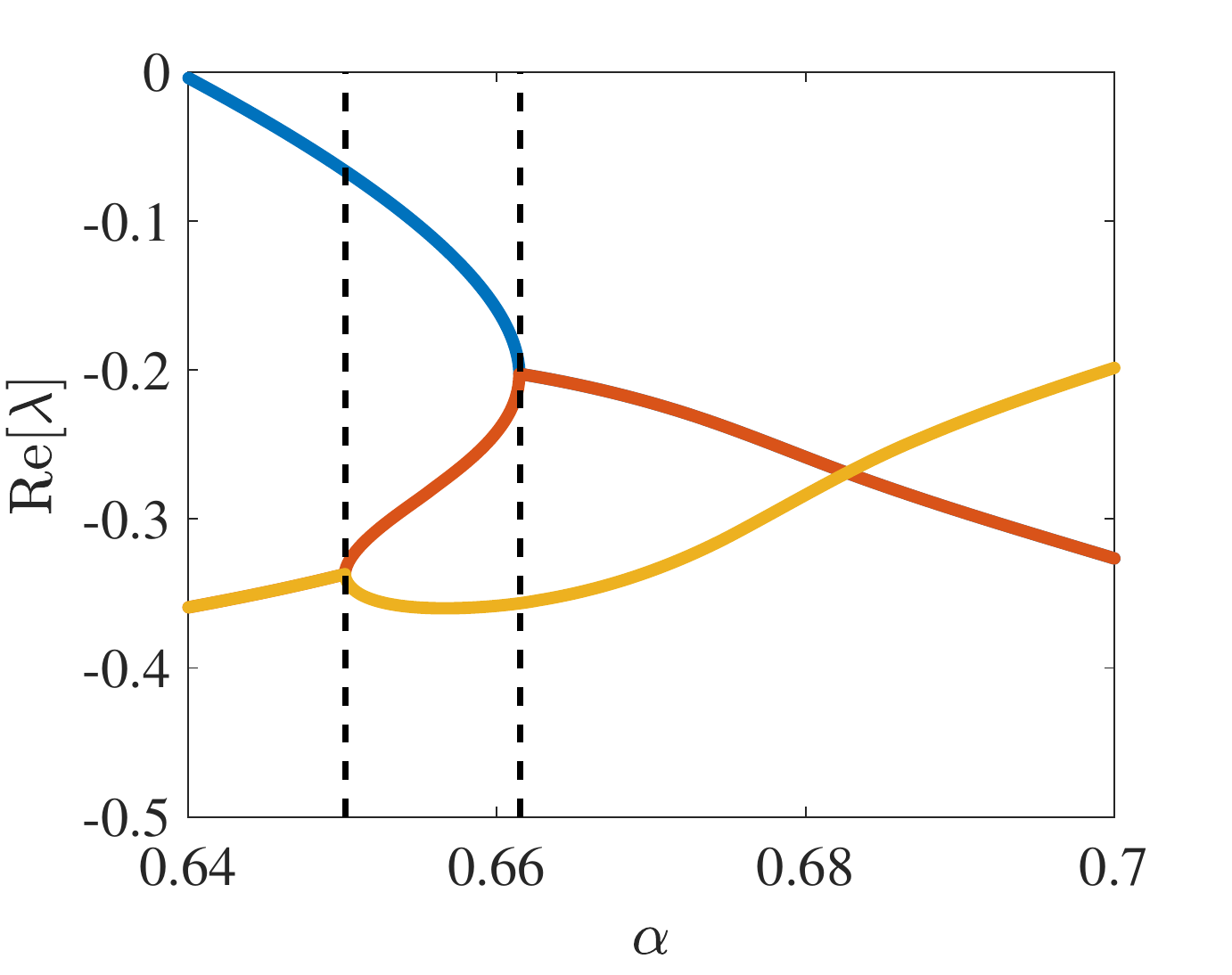}
   \label{fig:Fig7a}}
\subfigure[]{\includegraphics[width=4.1cm]{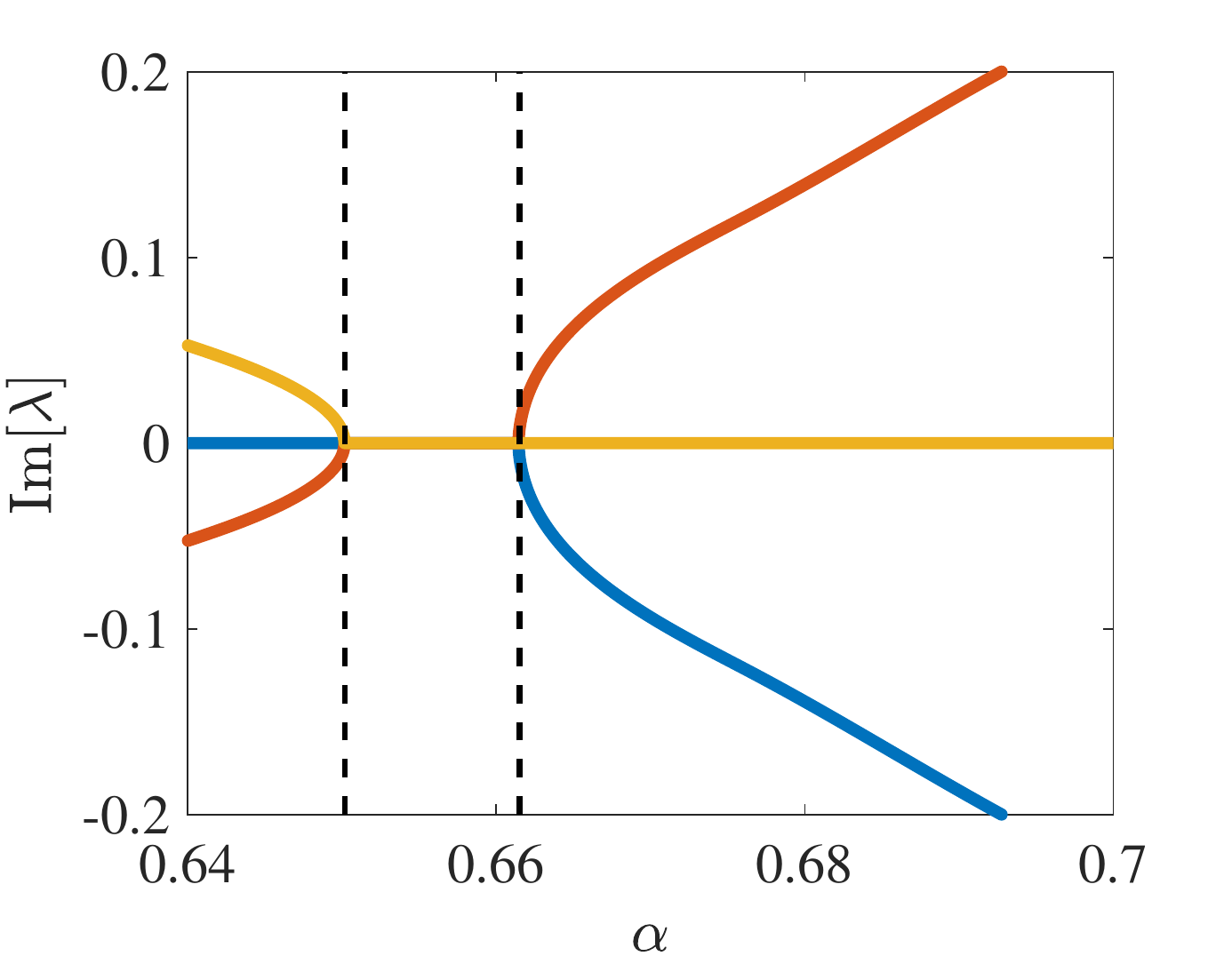}
   \label{fig:Fig7b}}\\
\subfigure[]{\includegraphics[width=4.1cm]{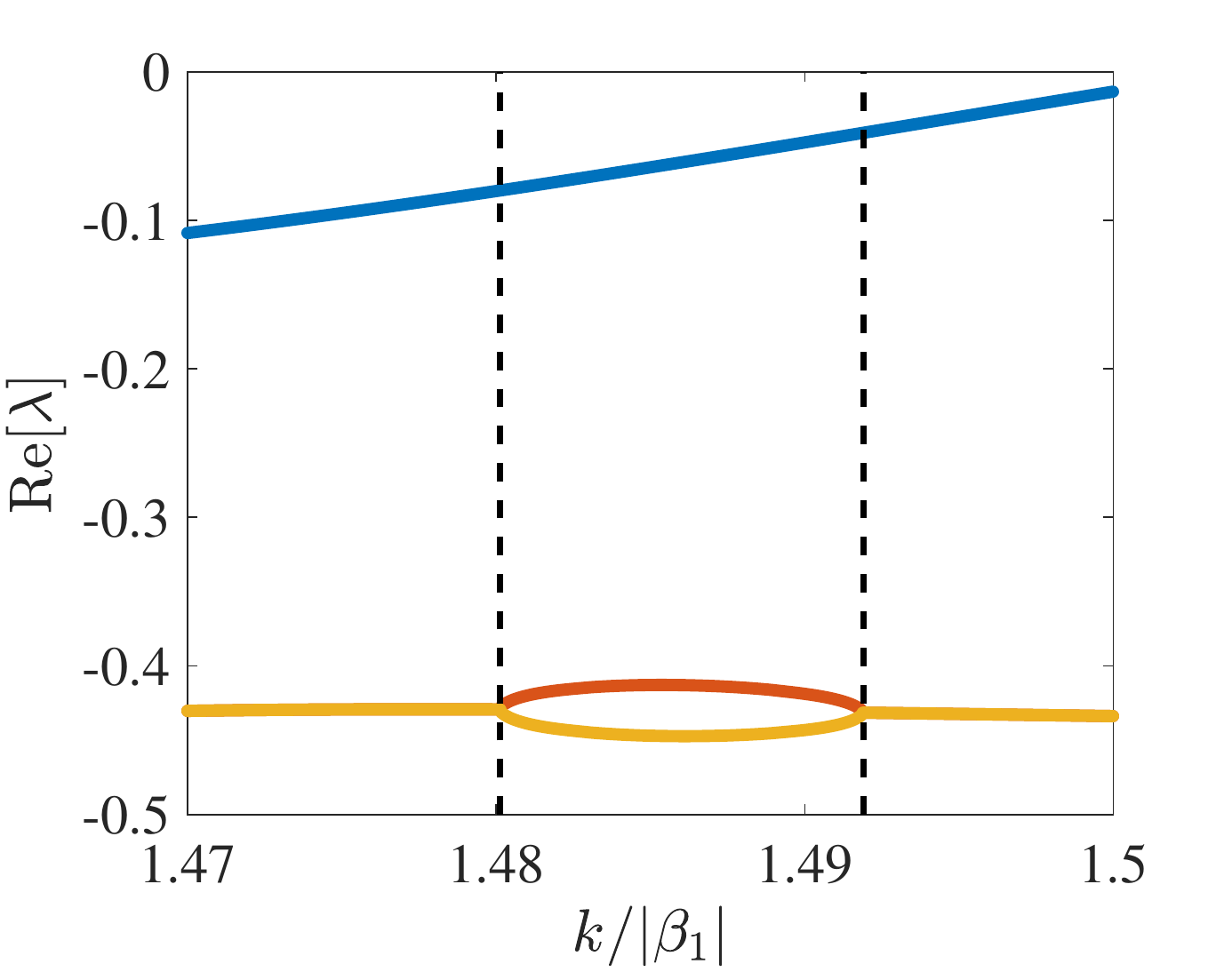}
   \label{fig:Fig8a}}
\subfigure[]{\includegraphics[width=4.1cm]{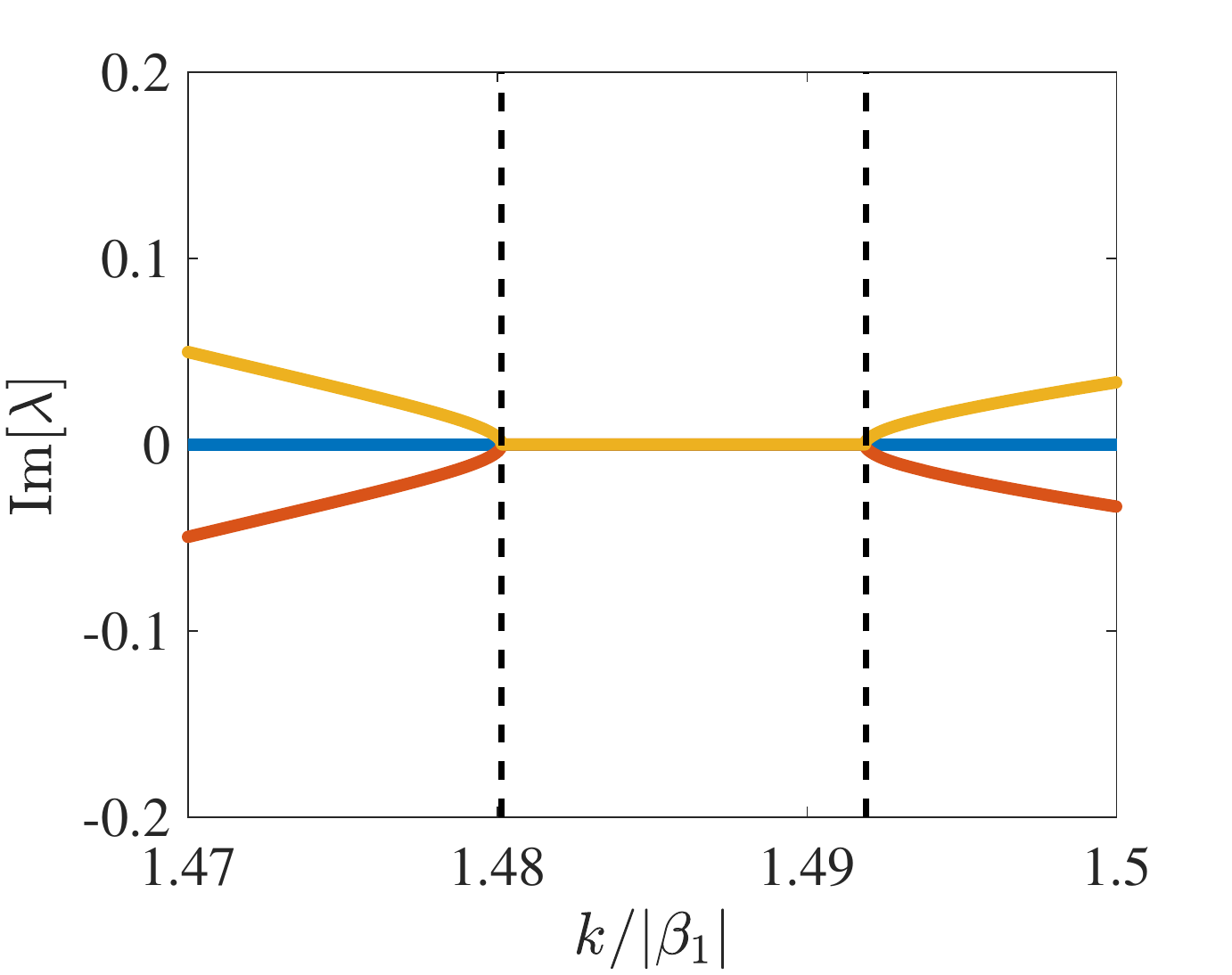}
   \label{fig:Fig8b}}
\caption{The: (a) real and (b) imaginary part of the eigenvalues as function of gain/loss contrast $\alpha$ for $k/|\beta_1|=1.6$. The: (a) real and (b) imaginary part of the eigenvalues as a function of the coupling $k/|\beta_1|$ for $\alpha=0.565$. Vertical dashed lines denote exceptional points. Common plot parameters: $\beta=1.012$, $\e=0.8$ and $\beta_1=-1$.}
\label{fig:Figs78}
\end{figure}

\subsection{Eigenvalue Spectra and Exceptional Points}
The eigenvalues $\lambda$ of the Jacobian of the linearized system which determine its behavior around a mode of operation can coalesce for specific selections of parameters. Such a spectral degeneracy is typical for non-Hermitian systems (with gain and loss) and is known as an Exceptional Point (EP); in its vicinity, the behavior of the system is of fundamentally different nature compared to the neighboring points \cite{AluMiri, Moiseyev_08}. This spectral coincidence offers opportunities for ultrasensitive measurements \cite{Christodoulides, Yang_17} and other interesting applications involving energy transfer between developed waves \cite{Vahala, Chong_16}. Moreover, the EP have been shown to possess specific spectral signatures in the noise and modulation response \cite{Kominis_17c} of non-Hermitian dimers and occur in abundance at the generic configuration of a dimer, not restricted by any symmetry conditions or other requirements related to the optical frequency mismatch and gain/loss ratio \cite{Kominis_18}. It is, therefore, meaningful to examine the occurrence of EPs in the system investigated in the present report. We have performed a thorough search across the considered parameter space in the quest of coalescing eigenvalues $\lambda$ and determined certain parametric loci along which such a degeneracy occurs into stable regions. 


In Figs \ref{fig:Fig7a} and \ref{fig:Fig7b}, we show the variation of the real and imaginary parts of the three eigenvalues $\lambda$ as function of gain/loss contrast $\alpha$; the black dashed vertical lines denote the emergence of EPs. Starting from Fig. \ref{fig:Fig7b}, we observe that two EPs appear in pairs at those $\alpha$'s that all the three eigenvalues $\lambda$'s convert from complex into real and vice-versa; such a result is anticipated since the coefficients of the characteristic polynomial $|\textbf{J}-\lambda\textbf{I}|$ (with $\textbf{J}$ given by Eq. (\ref{Jacobian}) and $\textbf{I}$ being the $3 \times 3$ identity matrix) are real. All these transitions happen within a restricted $\alpha$-range where the system is stable as indicated by Fig. \ref{fig:Fig7a} ($\Re[\lambda]<0$ for all eigenvalues). In Figs \ref{fig:Fig8a} and \ref{fig:Fig8b}, we show the dependence of the real and imaginary parts of $\lambda$'s respectively as functions of coupling coefficient $k/|\beta_1|$. It seems that, by adjusting the distance between the two waveguides, one can achieve twice the coalescence of eigenvalues. 

\begin{figure}[ht!]
\centering
\subfigure[]{\includegraphics[width=4.1cm]{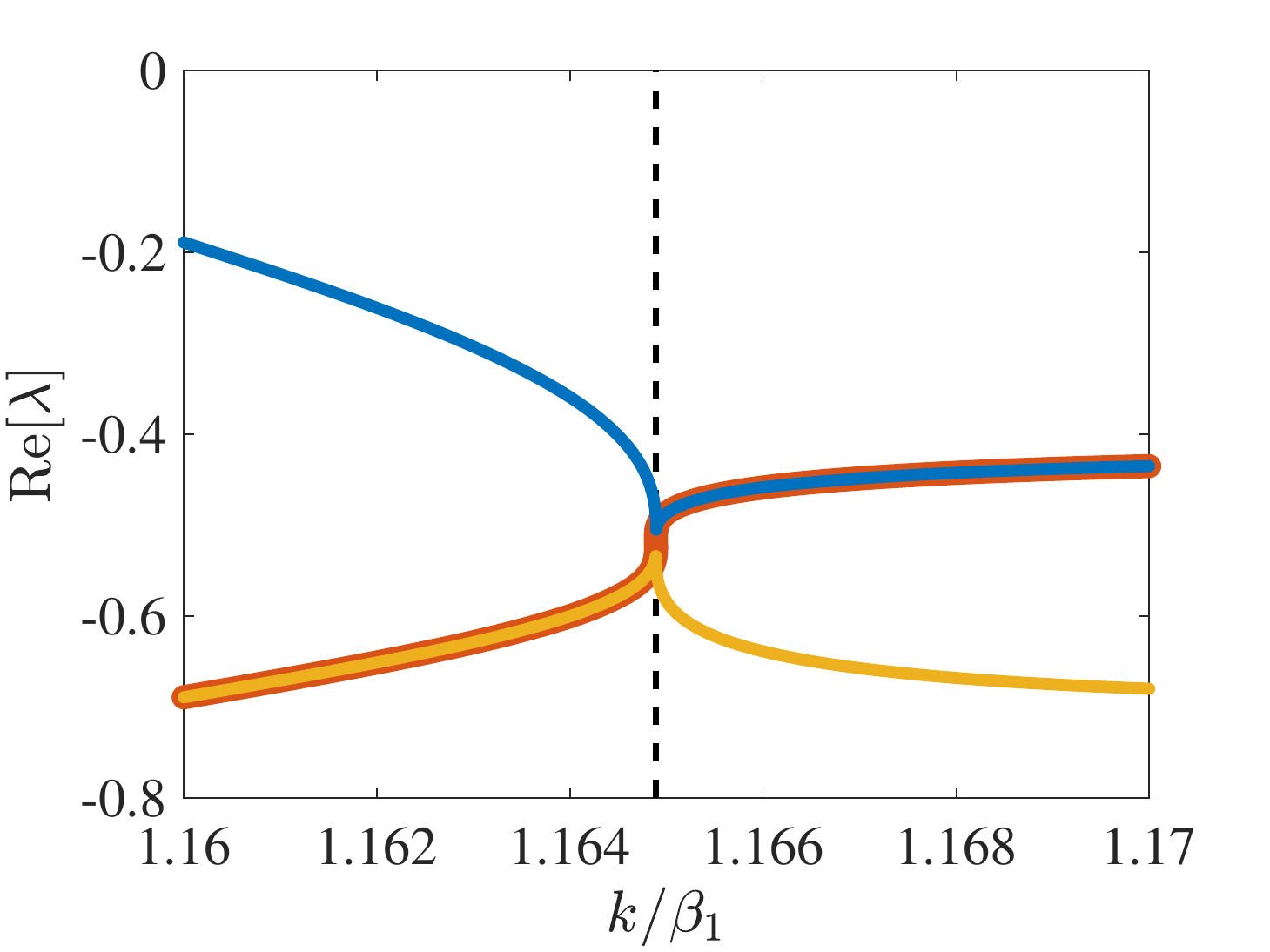}
   \label{fig:Fig10a}}
\subfigure[]{\includegraphics[width=4.1cm]{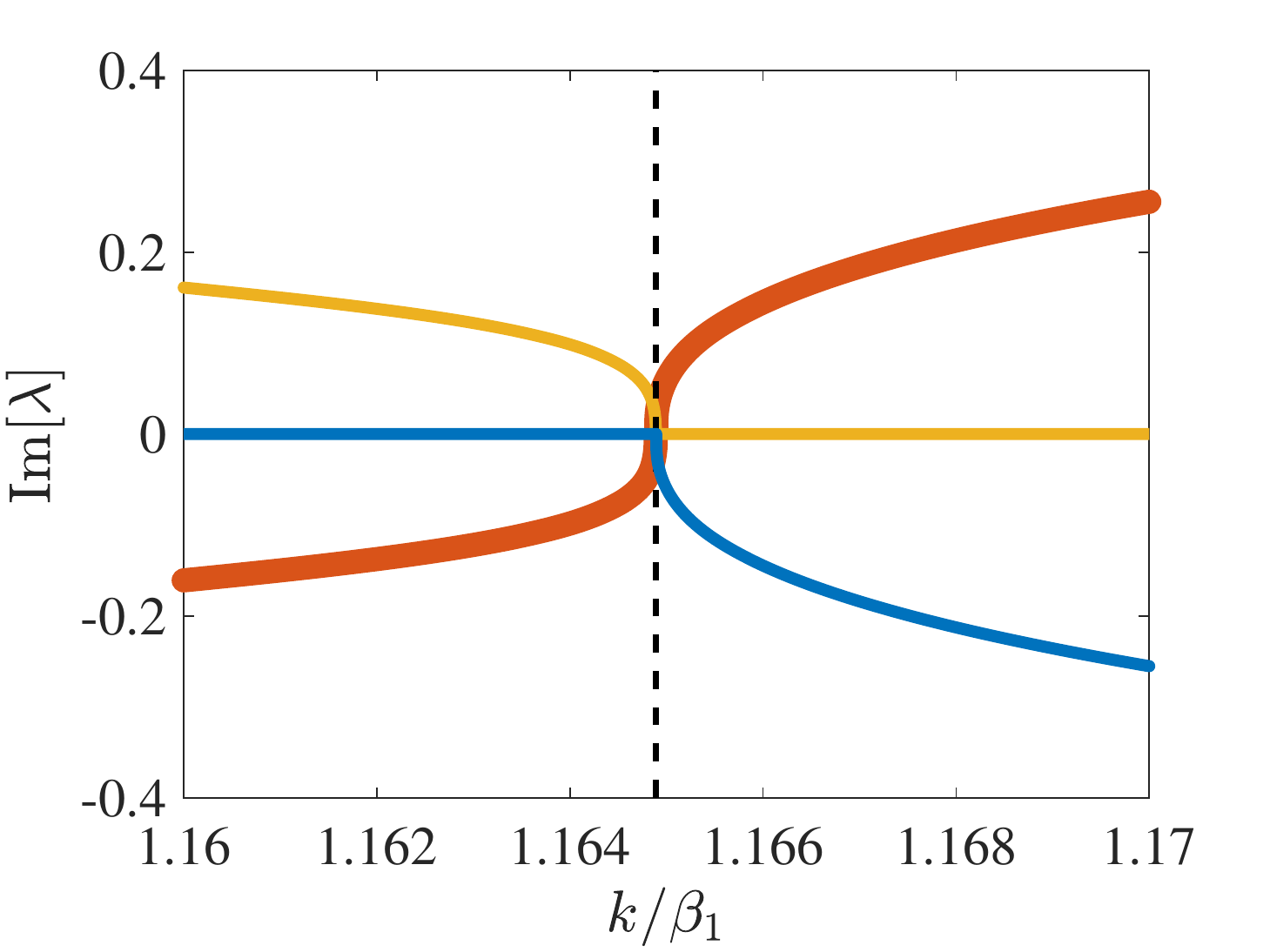}
   \label{fig:Fig10b}}
\caption{The: (a) real and (b) imaginary part of the eigenvalues as function of the coupling $k/\beta_1$ for $\alpha=0.521$. Vertical dashed line denote EPs. Plot parameters: $\beta=0.3$, $\e=0.5$ and $\beta_1=1$.}
\label{fig:Figs10}
\end{figure}

Finally, in Figs \ref{fig:Figs10}, we show the case of an EP occurring at the coalescence of all three of the system's eigenvalues $\lambda$ as the coupling $k$ is swept for $\beta_1>0$. Two of the eigenvalues follow opposite trends in their real part and meet the third one at a specific $k$; as far as the imaginary parts are concerned, they are zero only at the EP, unlike the ones in Figs \ref{fig:Fig7b} and \ref{fig:Fig8b}. We notice again that a careful scanning of the coupling strength can lead to highly effective exceptional point-based designs like polarization converters \cite{DynamicallyEncircling} or optical amplifiers \cite{EPBasedOPAM}.

\section{Concluding Remarks}
The most generic configuration of an asymmetric active photonic coupler with saturable gain has been studied on the basis of coupled-mode equations valid for any non-Hermitian dimer. The saturation has been shown to result in significant stability enhancement of the nonlinear supermodes of the system, as well as avoidance of undesirable evolution to blow-up solutions. The extent of the stability domains in parameter space, as well as the basins of attraction of the stable nonlinear supermodes in the phase space of the system were studied in detail. The critical role of asymmetry between the active and the passive waveguide parameters is demonstrated in terms of the number of the supported nonlinear supermodes and the presence of bistability. Moreover, the existence of exceptional points in the linear spectrum of the stable modes has been shown to extensively occur in the parameter space of the system, namely the spectral degeneracies are accessible by varying any one of the key parameters. All these important features suggest that the asymmetric active photonic coupler is a fundamental element capable of reconfigurable functionality for photonic integrated circuits applications spanning from metrology and optical sensors to phase switching and ultrafast communications. Finally, we believe that the systematic investigation presented in this manuscript can be used to reveal the full complexity of the basins of attraction, not only of stable steady states, but also of other coexisting attractors, such as periodic oscillations (limit cycles), in optically injected photonic oscillators and systems of two optically coupled semiconductor lasers, where the asymmetry and the nonlinear gain coefficient are seen to play a fundamental role.

\begin{acknowledgments}
This work has been supported by Nazarbayev University with Small Grant No. 090118FD5349 (``Super transmitters, radiators and lenses via photonic synthetic matter'') and an ORAU grant 2017-2020 on "Taming Chimeras to Achieve the Superradiant Emitter". Funding from MES RK state-targeted program BR05236454 is also acknowledged.
\end{acknowledgments}


\begin{thebibliography}{100}

\bibitem{Jensen_82} 
S.M. Jensen, 
``The Nonlinear Coherent Coupler,'' 
IEEE J. Quant. Electron. \textbf{18}, 1580-1583 (1982).

\bibitem{Daino_85} 
B. Daino, G. Gregori, and S. Wabnitz, 
``Stability analysis of nonlinear coherent coupling,'' 
J. Appl. Phys. \textbf{58}, 4512-4514 (1985).

\bibitem{Saleh} 
B. E. A. Saleh and M. C. Teich, 
\textit{Fundamentals of Photonics} (Wiley, 2007).

\bibitem{Feng_13} 
L.Feng, Y.-L. Xu, W. S. Fegadolli, M.-H. Lu, J. E. B. Oliveira, V. R. Almeida, Y.-F. Chen, and A. Scherer, 
``Experimental demonstration of a unidirectional reflectionless parity-time metamaterial at optical frequencies,'' 
Nat. Mater. \textbf{12}, 108-113 (2013).

\bibitem{Feng_11} 
L. Feng, M. Ayache, J. Huang, Y.-L. Xu, M.-H. Lu, Y.-F. Chen, Y. Fainman, and A. Scherer, 
``Nonreciprocal Light Propagation in a Silicon Photonic Circuit,'' 
Science \textbf{333}, 729-733 (2011).

\bibitem{Jalas_13} 
D. Jalas \textit{et al.}, 
``What is - and what is not - an optical isolator,'' 
Nature Photon. \textbf{7}, 579-582 (2013).

\bibitem{Shi_15} 
Y. Shi, Z. Yu, and S. Fan, 
``Limitations of nonlinear optical isolators due to dynamic reciprocity,'' 
Nature Photon. \textbf{9}, 388-392 (2015).

\bibitem{Thompson_86} 
G. H. B. Thompson, 
``Analysis of Optical Directional Couplers That Include Gain or Loss and Their Application to Semiconductor Slab Dielectric Guides,'' 
J. Light. Technol. \textbf{4}, 1678-1693 (1986).

\bibitem{Chen_92} 
Y. Chen, A. W. Snyder, and D. N. Payne, 
``Twin Core Nonlinear Couplers with Gain and Loss,'' 
IEEE J. Quant. Electron. \textbf{28}, 239-245 (1992).

\bibitem{Malomed_17} 
J. D. Huerta Morales, B. M. Rodriguez-Lara, and B. A. Malomed, 
``Polarization dynamics in twisted fiber amplifiers: a non-Hermitian nonlinear dimer model,'' 
Opt. Lett., \textbf{42}, 4402-4405 (2017).

\bibitem{Kenkre_86}  
V. M. Kenkre and D. K. Campbell, 
``Self-trapping on a dimer: Time-dependent solutions of a discrete nonlinear Schrodinger equation,'' 
Phys. Rev. B, \textbf{34}, 4959-4961 (1986).

\bibitem{Dattoli_90} 
G. Dattoli, A. Torre, and R. Mignani, 
``Non-Hermitian evolution of two-level quantum systems,'' 
Phys. Rev. A \textbf{42}, 1467-1475 (1990).


\bibitem{DawnNonHermitian}
R. El-Ganainy, M. Khajavikhan, D. N. Christodoulides, and S. K. Ozdemir,
``The dawn of non-Hermitian optics,''
Commun. Phys. \textbf{2}, 37 (2019).

\bibitem{PTWaveguides}
H. Alaeian and J. A. Dionne,
``Non-Hermitian nanophotonic and plasmonic waveguides,''
Phys. Rev. B \textbf{89}, 075136 (2014).

\bibitem{MyJ73}
F. Monticone, C. A. Valagiannopoulos, and A. Alù,
``Parity-Time Symmetric Nonlocal Metasurfaces: All-Angle Negative Refraction and Volumetric Imaging,''
Phys. Rev. X \textbf{6}, 041018 (2016).

\bibitem{PTPlasmonics}
C. A. Valagiannopoulos,
``Optical PT-Symmetric Counterparts of Ordinary Metals,''
IEEE J. Sel. Top. Quantum Electron. \textbf{22}, 5000409 (2016).

\bibitem{BenistySwitches} 
A. Lupu, H. Benisty, and A. Degiron,
``Switching using PT symmetry in plasmonic systems: positive role of the losses,''
Opt. Express \textbf{21}, 21651-21668 (2013).

\bibitem{InvisibleAcousticSensor}
R. Fleury, D. Sounas, and A. Alù,
``An invisible acoustic sensor based on parity-time symmetry,''
Nat. Commun. \textbf{6}, 5905 (2015).

\bibitem{SingleModeLaser} 
L. Feng, Z. J. Wong, R.-M. Ma, Y. Wang, X. Zhang,
``Single-mode laser by parity-time symmetry breaking,''
Science \textbf{346}, 972-975 (2014).

\bibitem{PTSymmetricMicrorings} 
H. Hodaei, M.-A. Miri, M. Heinrich, D. N. Christodoulides, and M. Khajavikhan,
``Parity-time-symmetric microring lasers,''
Science \textbf{346}, 975-978 (2014).

\bibitem{Longhi_10}
S. Longhi,
``PT-symmetric laser absorber,'' 
Phys. Rev. A. \textbf{82}, 031801 (2010)

\bibitem{NonHermitianPhotonics} 
L. Feng, R. El-Ganainy, and L. Ge,
``Non-Hermitian photonics based on parity–time symmetry,''
Nat. Photonics \textbf{11}, 752-762 (2017).

\bibitem{WhisperingGallery} 
B. Peng, S. K. Özdemir, F. Lei, F. Monifi, M. Gianfreda, G. L. Long, S. Fan, F. Nori, C. M. Bender, and L. Yang,
``Parity–time-symmetric whispering-gallery microcavities,''
Nat. Physics \textbf{10}, 394-398 (2014).

\bibitem{NonlinearSwitching}
S. V. Suchkov, A. A. Sukhorukov, J. Huang, S. V. Dmitriev, C. Lee, and Y. S. Kivshar,
``Nonlinear switching and solitons in PT-symmetric photonic systems,''
Laser Photonics Rev. \textbf{10}, 177–213 (2016).

\bibitem{Ramezani_10} 
H. Ramezani, T. Kottos, R. El-Ganainy, and D. N. Christodoulides, 
``Unidirectional nonlinear PT-symmetric optical structures,'' 
Phys. Rev. A \textbf{82}, 043803 (2010).

\bibitem{NonlinearWaves}
V. V. Konotop, J. Yang, and D. A. Zezyulin,
``Nonlinear waves in PT-symmetric systems,''
Rev. Mod. Phys. \textbf{88}, 035002 (2016).


\bibitem{Barashenkov_13} 
I. V. Barashenkov, G. S. Jackson and S. Flach, 
``Blow-up regimes in the PT-symmetric coupler and the actively coupled dimer,'' 
Phys. Rev. A \textbf{88}, 053817 (2013).

\bibitem{Christodoulides_16} 
A. U. Hassan, H. Hodaei, M.-A. Miri, M. Khajavikhan, and D. N. Christodoulides,
``Integrable nonlinear parity-time-symmetric optical oscillator,'' 
Phys. Rev. E \textbf{93}, 042219 (2016).

\bibitem{Kominis_17} 
Y. Kominis, T. Bountis, and S. Flach, 
``Stability through asymmetry: Modulationally stable nonlinear supermodes of asymmetric non-Hermitian optical couplers,'' 
Phys. Rev. A \textbf{95}, 063832 (2017).

\bibitem{Kominis_16} 
Y. Kominis, T. Bountis, and S. Flach, 
``The Asymmetric Active Coupler: Stable Nonlinear Supermodes and Directed Transport,'' 
Sci. Rep. \textbf{6}, 33699 (2016).

\bibitem{Kominis_17b} 
Y. Kominis, V. Kovanis, T. Bountis, 
``Controllable Asymmetric Phase-Locked States of the Fundamental Active Photonic Dimer,'' 
Phys. Rev. A \textbf{96}, 043836 (2017).

\bibitem{Choquette_17} 
Z. Gao, S.T.M Fryslie, B.J. Thompson, P. Scott Carney, and K.D. Choquette, 
``Parity-time symmetry in coherently coupled vertical cavity laser arrays,'' 
Optica \textbf{4}, 323-329 (2017).

\bibitem{Roy_19} 
J.D. Hart, Y. Zhang, R. Roy, and A.E. Motter, 
``Topological Control of Synchronization Patterns: Trading Symmetry for Stability,'' 
Phys. Rev. Lett. \textbf{122}, 058301 (2019).

\bibitem{Kominis_19b} 
Y. Kominis, J. Cuevas-Maraver, P. G. Kevrekidis, D. J. Frantzeskakis, and A. Bountis, 
``Continuous families of solitary waves in non-symmetric complex potentials: A Melnikov theory approach,'' 
Chaos Solitons Fract. \textbf{118} 222–232 (2019). 

\bibitem {Kominis_15a} Y. Kominis, 
``Soliton dynamics in symmetric and non-symmetric complex potentials,'' 
Opt. Commun. \textbf{334}, 265–272 (2015).

\bibitem{Kominis_15b} 
Y. Kominis, 
``Dynamic power balance for nonlinear waves in unbalanced gain and loss landscapes,'' 
Phys. Rev. A \textbf{92}, 063849 (2015).


\bibitem{NonlinearGain}
D. T. Nichols and H. G. Winful, 
``The effect of nonlinear gain on the stability of evanescently coupled semiconductor laser arrays,'' 
J. Appl. Phys. \textbf{73}, 459-461 (1993).


\bibitem{OpticalCavityEffects}
J. C. Johnson, H. Yan, P. Yang, and R. J. Saykally, 
``Optical Cavity Effects in ZnO Nanowire Lasers and Waveguides,'' 
J. Phys. Chem. B \textbf{107}, 8816-8828 (2003).

\bibitem{UltrafastPhotonicCrystal}
H. Nakamura \textit{et al.}, 
``Ultra-fast photonic crystal/quantum dot all-optical switch for future photonic networks,'' 
Opt. Express \textbf{12}, 6606-6614 (2004).


\bibitem{Thirstrup_95} 
C. Thirstrup, 
``Optical Bistability in a Nonlinear Directional Coupler,'' 
IEEE J. Quant. Electron. \textbf{31}, 2101-2106 (1995).

\bibitem{FilmCoupledMetasurfaces} 
Z. Huang, A. Baron, S. Larouche, C. Argyropoulos, and D. R. Smith,
``Optical bistability with film-coupled metasurfaces,''
Opt. Lett. \textbf{40}, 5638-5641 (2015).

\bibitem{Ludge_18}
A. Rohm, K. Ludge, and I. Schneider, 
``Bistability in two simple symmetrically coupled oscillators with symmetry-broken amplitude- and phase-locking,'' 
Chaos \textbf{28}, 063114 (2018).


\bibitem{Thompson} 
J. M. T. Thompson and H. B. Stewart, 
\textit{Nonlinear Dynamics and Chaos} (Willey, 2002).

\bibitem{PeriodicOrbitsBasins}
I. Sliwa and K. Grygiel,
``Periodic orbits, basins of attraction and chaotic beats in two coupled Kerr oscillators,''
Nonlinear Dyn. \textbf{67}, 755-765 (2012).

\bibitem{Roukes_07} 
I. Kozinsky, H.W.Ch. Postma, O. Kogan, A. Husain, and M.L. Roukes, 
``Basins of Attraction of a Nonlinear Nanomechanical Resonator,'' 
Phys. Rev. Lett. \textbf{99}, 207201 (2007).


\bibitem{Heiss} 
W. D. Heiss, 
``The physics of exceptional points,'' 
J. Phys. A: Math. Theor. \textbf{45}, 444016 (2012). 

\bibitem{Christodoulides}
H. Hodaei, A. U. Hassan, S. Wittek, H. Garcia-Gracia, R. El-Ganainy, D. N. Christodoulides, and M. Khajavikhan,
``Enhanced sensitivity at higher-order exceptional points,''
Nature \textbf{548}, 187-191 (2017).

\bibitem{Fan_19} 
H. Wang, S. Assawaworrarit, and S. Fan, 
``Dynamics for encircling an exceptional point in a nonlinear non-Hermitian system,'' 
Opt. Lett. \textbf{44}, 638-641 (2019).


\bibitem{CMT}
W.-P. Huang,
``Coupled-mode theory for optical waveguides: an overview,''
J. Opt. Soc. Am. A \textbf{11}, 963-983 (1994).








\bibitem{CascadingElasticVibrations}
O. R. Bilal, A. Foehr, and C. Daraio,
``Bistable metamaterial for switching and cascading elastic vibrations,''
Proc. Natl. Acad. Sci. U.S.A. \textbf{18}, 114 (2017).

\bibitem{ThermalMemory} 
V. Kubytskyi, S.-A. Biehs, P. Ben-Abdallah,
``Radiative Bistability and Thermal Memory,''
Phys. Rev. Lett. \textbf{113}, 074301 (2014).

\bibitem{InMemory} 
C. Ríos, N. Youngblood, Z. Cheng, M. Le Gallo, W. H. P. Pernice, C. D. Wright, A. Sebastian, and H. Bhaskaran,
``In-memory computing on a photonic platform,''
Sci. Adv. \textbf{5}, eaau5759 (2019).

\bibitem{AtomLightInteractions} 
A. Goban, C.-L. Hung, S.-P. Yu, J. D. Hood, J. A. Muniz, J. H. Lee, M. J. Martin, A. C. McClung, K. S. Choi, D. E. Chang, O. Painter, and H. J. Kimble,
``Atom–light interactions in photonic crystals,''
Nat. Commun. \textbf{5}, 3808 (2014).




%

\bibitem{ComplicatedBasins}
A. Prasad, Y.-C. Lai, A. Gavrielides, and V. Kovanis,
``Complicated basins in external-cavity semiconductor lasers,''
Phys. Lett. A \textbf{314}, 44-50 (2003).


\bibitem{AluMiri}
M.-A. Miri and A. Alù,
``Exceptional points in optics and photonics,''
Science \textbf{363}, eaar7709 (2019).

\bibitem{Moiseyev_08} 
S. Klaiman, U. Gunther, and N. Moiseyev, 
``Visualization of Branch Points in PT-Symmetric Waveguides,'' 
Phys. Rev. Lett. \textbf{101}, 080402 (2008).

\bibitem{Yang_17} 
W. Chen, S. K. Ozdemir, G. Zhao, J. Wiersig, and L. Yang, 
``Exceptional points enhance sensing in an optical microcavity,'' 
Nature \textbf{548}, 192195 (2017).

\bibitem{Vahala}
Y.-H. Lai, Y.-K. Lu, M.-G. Suh, K. Vahala,
``Enhanced sensitivity operation of an optical gyroscope near an exceptional point,''
Submitted to ArXiv, \url{https://arxiv.org/abs/1901.08217} (2019).

\bibitem{Chong_16} 
S. N. Ghosh and Y.D. Chong, 
``Exceptional points and asymmetric mode conversion in quasi-guided dual-mode optical waveguides,'' 
Sci. Rep. \textbf{6}, 19837 (2016).

\bibitem{Kominis_17c} 
Y. Kominis, V. Kovanis, T. Bountis, 
``Spectral Signatures of Exceptional Points and Bifurcations in the Fundamental Active Photonic Dimer,'' 
Phys. Rev. A \textbf{96}, 053837 (2017).

\bibitem{Kominis_18} 
Y. Kominis, K. D. Choquette, T. Bountis, and V. Kovanis, 
``Exceptional Points in Two Dissimilar Coupled Diode Lasers,'' 
Appl. Phys. Lett. \textbf{113}, 081103 (2018).

\bibitem{DynamicallyEncircling}
A. U. Hassan, B. Zhen, M. Soljacic, M. Khajavikhan, and D. N. Christodoulides,
``Dynamically Encircling Exceptional Points: Exact Evolution and Polarization State Conversion,''
Phys. Rev. Lett. \textbf{118}, 093002 (2017).

\bibitem{EPBasedOPAM}
Q. Zhong, S. K. Ozdemir, A. Eisfeld, A. Metelmann, and R. El-Ganainy,
``Exceptional points-based optical amplifiers,'' 
arXiv, \url{https://arxiv.org/abs/1904.13005} (2019).

\end{thebibliography}
\end{document}